\documentclass[12pt,reprint,aps,
  pra,
  amsmath,amssymb,longbibliography]{revtex4-1}
\usepackage[varg]{txfonts}
\usepackage{bm}
\usepackage{float}
\usepackage{xcolor}
\usepackage[caption=false]{subfig}
\PassOptionsToPackage{hyphens}{url}
\usepackage[bookmarks=true,         
     pdfnewwindow=true,      
     colorlinks=true,    
     linkcolor=blue,     
     citecolor=blue,     
     filecolor=blue,  
     urlcolor=blue,      
     final=true,
 ]{hyperref}
\usepackage{color,epsfig,amsmath,amssymb,fancyhdr} 

\usepackage{footnote}
\usepackage{lipsum}

\usepackage{booktabs}

\begin{document}

\title{Penetration of a cooling convective layer into a stably-stratified composition gradient: entrainment at low Prandtl number}

\author{J. R. Fuentes}
\author{A. Cumming}

\affiliation{Department of Physics and McGill Space Institute, McGill University, 3600 rue University, Montreal, QC H3A 2T8, Canada}

\begin{abstract}

We study the formation and inward propagation of a convective layer when a stably-stratified fluid with a composition gradient is cooled from above. We perform a series of two-dimensional simulations using the Bousinessq approximation with Prandtl number ranging from $Pr = 0.1$ to $7$, extending previous work on salty water to low $Pr$. We show that the evolution of the convection zone is well-described by an entrainment prescription in which a fixed fraction of the kinetic energy of convective motions is used to mix fluid at the interface with the stable layer. We measure the entrainment efficiency and find that it grows with decreasing Prandtl number or increased applied heat flux. The kinetic energy flux that determines the entrainment rate is a small fraction of the thermal energy flux carried by convective motions. In this time-dependent situation, the density ratio at the interface is driven to a narrow range that depends on the value of $Pr$, and with low enough values that advection dominates the interfacial transport. We characterize the interfacial flux ratio and how it depends on the interface stability. We present an analytic model that accounts for the growth of the convective layer with two parameters, the entrainment efficiency and the interfacial heat transport, both of which can be measured from the simulations.
\end{abstract}

\maketitle

\section{Introduction}

In astrophysics, there are many situations in which a convective zone coexists with a composition gradient. A classic example is the convective core of a massive star, which is immersed in a gradient of heavy elements that results from nuclear burning \citep{1947ApJ...105..305L,1954ApJ...120..332T,1958ApJ...128..348S,1995ApJ...444..318M}. Gas giant planets, which undergo strong convection in their gaseous envelopes, might develop composition gradients from either their formation history, or collisions during their evolution \citep{2007ApJ...661L..81C,2012A&A...540A..20L}. Recently, the Juno mission \citep{2017SSRv..213....5B} has found evidence that an extended region of Jupiter's interior is enriched in heavy elements \citep{2017GeoRL..44.4649W,2019ApJ...872..100D}. 

The nature of convective mixing in these regions is not clear.  It is well known that composition gradients tend to stabilize the fluid against overturning convection \citep{1966PASJ...18..374K}, but the resulting transport of heat and heavy elements is not well-understood. In stellar evolution, mixing across the boundary between a convection zone and a stable region can be extremely important because it can bring fresh fuel for nuclear reactions into the convection zone. Evolution models for astrophysical objects over long timescales rely on analytic prescriptions for transport both within the convection zone and at the boundary. These are typically based on mixing length theory \citep[e.g.,][]{1983A&A...126..207L,2013A&A...552A..76S} and then implemented into one-dimensional numerical models \citep[e.g.,][]{2011ApJS..192....3P}.

Observations and numerical simulations of geophysical fluids with composition gradients have shown that under certain circumstances, double-diffusive instabilities lead to a series of convective layers. The layers are well-mixed in both composition and temperature, but separated by sharp interfaces across which transport of heat and composition is by molecular diffusion \citep[e.g.,][and references therein]{2003JFM...497..365R}. Astrophysical fluids differ in a key aspect, that the Prandtl number $Pr=\nu/\kappa_T$, which measures the ratio of kinematic viscosity $\nu$ to thermal diffusivity $\kappa_T$, is $Pr<1$ as opposed to $Pr\approx 7$ for salty water. 
Recently, with the improvement of computational resources, three-dimensional numerical simulations at low Prandtl numbers appropriate for planetary interiors ($Pr=\nu/\kappa_T\sim 0.001$--$0.1$) have become possible. This work shows that while thermo-composional convective layers can also exist at low $Pr$ \citep{2011ApJ...731...66R,2012ApJ...750...61M,2013ApJ...768..157W,2016ApJ...823...33M}, there are fundamental differences in how and whether layers form and the nature of doubly-diffusive convection \citep{2016ApJ...823...33M}. The reader is referred to the excellent review by \citet{2018AnRFM..50..275G} for further details. These simulations have guided new transport prescriptions that can be included in 1D evolution codes \cite{2016ApJ...823...33M}. Conditions in stellar interiors, where $Pr\lesssim 10^{-6}$, are still inaccessible numerically.

Despite the progress in understanding layer formation when there are pre-existing temperature and composition gradients, less attention has been paid to situations in which the large-scale gradients develop over time. An example is the penetration of a convective region into a neighbouring stable region with a composition gradient. This configuration is relevant in the evolution of gas giant planets, in which a convective zone propagates inwards as the planet cools down, enriching its outer regions by transporting heavy elements from below \citep[e.g.,][]{2018A&A...610L..14V,2020arXiv200413534M}. In this context, there are two relevant questions: 1) how quickly does the outer convective layer move inwards, and 2) does the fluid become fully-mixed? In the context of Jupiter, for example, recent 1D evolutionary models find that global composition gradients can persist over long timescales, by separating into a number of distinct convective layers, although not over as extensive a region as inferred from the Juno data \cite{2018A&A...610L..14V,2020arXiv200413534M}. These simulations, however, lack a detailed model of how composition and heat are transported at convective boundaries.

Several laboratory studies have been carried out in which stably-stratified salty water is heated from below, creating a convective region that penetrates into the stably-stratified layer \citep{1964PNAS...52...49T,turner_1968, huppert_linden_1979,fernando_1987}. 
Motivated by experimental results, \citet{turner_1968} developed a simple analytical model for the growth of the convective layer.  The fluid is assumed to be initially isothermal with a linear salinity gradient $dS/dz < 0$, and a constant heat flux $F_0$ is applied at the bottom boundary. 
The model assumes that at the top of the well-mixed convection zone there is an abrupt step of both temperature $\Delta T$ and salinity $\Delta S$ (i.e., molecular diffusion of heat and salt are ignored). After a time $t$, when the convective zone has a thickness $h$, from heat and salinity balance it follows that
\begin{align}
&\rho_0 c_P \Delta T h = F_0 t\, , \label{eq_heat_conservation}\\
&\Delta S = \dfrac{1}{2}\left|\dfrac{dS}{dz}\right| h\, , \label{eq_salinity_conservation}
\end{align}
where $\rho_0$ is a background density and $c_P$ is the specific heat at constant pressure. The rate at which the convection zone grows depends on the stability of the interface, $\beta\Delta S/\alpha\Delta T\equiv R_\rho$, where $\beta$ and $\alpha$ are the coefficients of solute contraction and thermal expansion (both assumed to be positive constants). 
For a given value of $R_\rho$, eqs.~(\ref{eq_heat_conservation}) and (\ref{eq_salinity_conservation}) give
\begin{equation}
h(t) = \left(2R_\rho\right)^{1/2}\left(\frac{\alpha F_0}{\rho_0 c_P}\right)^{1/2}\left(\beta \left|\dfrac{dS}{dz}\right|\right)^{-1/2} t^{1/2}\, . \label{eq_h_general}
\end{equation}
\citet{turner_1968} considered two limits for $R_\rho$. One possibility is that the convection zone grows by Rayleigh-Taylor instabilities, when its temperature has increased enough to lower the density jump at the interface to $\Delta \rho \approx 0$, ie.~$R_\rho\approx 1$. However, additional mixing mechanisms could in principle transport heat and salt across a Rayleigh-Taylor stable interface, leading to a more rapid growth of the convective layer. For example, Kelvin-Helmholtz instabilities at the boundary can lead to entrainment of fluid from the stable layer \cite{fernando_1987}. As a limiting case, \citet{turner_1968} found $R_\rho=3$ under the assumption that the potential energy change from heating the convective layer is used to redistribute the heavy elements.

Both experimental and numerical results for salty water suggest that entrainment at the interface does in fact play a key role.
While the initial measurements of \citet{turner_1968} suggested that $R_\rho\approx 1$, later experiments by \citet{fernando_1987} showed that the density interface is stable with a non-negligible buoyancy jump across it ($R_\rho>1$ or $\Delta B \equiv - g \Delta \rho /\rho_0 < 0$). \citet{fernando_1987} proposed that mixing across the stable interface occurs due to shear motions near the interface, and predicted the same relation as in Eq.~(\ref{eq_h_general}) for the growth rate of the convective layer, but with $R_\rho$ replaced by a different constant that depends on the entrainment efficiency. In an attempt to clarify the discrepancy between \citet{turner_1968} and \citet{fernando_1987}, \citet{molemaker_dijkstra_1997} performed two-dimensional numerical simulations, with a similar set-up as in the classic laboratory experiments but cooled from above instead of heated from below. Their results agreed with \citet{fernando_1987}, giving support to entrainment as the mixing mechanism. They also found that diffusive heat flux through the interface is significant, modifying Eq.~\eqref{eq_h_general}. 

In this work, we investigate how low $Pr$ affects the growth of a convective layer into a composition gradient. While there has been some work done with a time-dependent background temperature profile at low $Pr$ \cite{2001PhDT.........8B,2019ThCFD..33..383Z}, it was focused on the formation and evolution of layers.
Here we focus on the physics behind the growth of the convective zone. In particular, we investigate the efficiency of entrainment at lower $Pr$ numbers. To accomplish this, we perform a series of two-dimensional numerical experiments of an incompressible fluid with a linear composition gradient, cooled from the top with a constant heat flux.
Our simulations were performed with $Pr$ ranging from 0.1 to 7 (i.e.~we also include the salty water regime for comparison), at fixed diffusivity ratio $\tau \equiv \kappa_S/\kappa_T=0.1$ (with $\kappa_S$ the solute diffusivity).

The paper is organised as follows. In Sect.~\ref{sec_model} we describe the physical model and the numerical code used to perform the simulations. Sect.~\ref{sec_inwards_prop} presents a description of the inwards propagation of the convective layer. In Sect.~\ref{sec_entrainment} we measure the entrainment efficiency at small $Pr$. In Sect.~\ref{sec_flux_below} we discuss the relevance of the heat flux across the interface between the convection zone and the stable layer, and its effect on the growth of the layer. In Sect.~\ref{sec_flux_transport} we discuss the relative sizes of heat and composition transport across the interface. In Sect.~\ref{sect_model} we present an analytic model of the evolution of the convective layer that reproduces our numerical results. Finally, we conclude in Sect.~\ref{sec_conclusions}.

\section{Model and numerical method} \label{sec_model}

We perform two-dimensional simulations in a horizontally-periodic domain of height $H$ and width $L$. We impose impermeable and stress-free top and bottom boundaries with no composition flux through them, no heat flux at the bottom, and a constant heat flux at the top. We use the Boussinesq approximation \citep{1960ApJ...131..442S}, valid for a thin layer of fluid in which fluctuations in density ($\rho$) are small with respect to the constant background density ($\rho_0$). The density variations depend on temperature and solute perturbations ($T$ and $S$, respectively) via $\rho = \rho_0 (\beta S - \alpha T)$, where $\beta$ and $\alpha$ are the coefficients of solute contraction and thermal expansion, respectively, both assumed to be positive constants. The governing equations are 
\begin{gather}
\nabla \cdot \bm{v} = 0\, ,\\
\dfrac{\partial T}{\partial t}  = - (\bm{v}\cdot \nabla)\,T + \kappa_T\nabla^2 T\, ,\\
\dfrac{\partial S}{\partial t} = - (\bm{v}\cdot \nabla)\,S + \kappa_S\nabla^2 S\, ,\\
\dfrac{\partial \bm{v}}{\partial t} = - (\bm{v}\cdot \nabla)\,\bm{v} - \dfrac{\nabla P}{\rho_0} + \left(\dfrac{\rho}{\rho_0}\right)\bm{g} + \nu\nabla^2 \bm{v}\,,
\end{gather}
with boundary conditions
\begin{gather}
w\,\big\vert_{z=0,H} = 0\,, \hspace{0.2cm} \dfrac{\partial  u}{\partial z}\,\bigg\vert_{z=0,H} = 0\,, \hspace{0.2cm} \dfrac{\partial S}{\partial z}\,\bigg\vert_{z=0,H} = 0\,,\\
\dfrac{\partial T}{\partial z}\,\bigg\vert_{z=0} = 0\, , \hspace{0.2cm} \dfrac{\partial T}{\partial z}\,\bigg\vert_{z=H} = -\dfrac{F_0}{k}\, .
\end{gather}
In the above equations, $\bm{v} = (u,w)$ is the velocity of a fluid element, where $u$ is the $x$-component, and $w$ is the $z$-component, $P$ denotes the pressure fluctuation resulting from the motion of the fluid, $\bm{g}$ is the acceleration due to gravity, and $k = \rho_0 c_P \kappa_T$ is the thermal conductivity. Further, $F_0$ corresponds to the constant heat flux at the top boundary that cools the domain.

The fluid is initialized with constant temperature $T_0$ everywhere and with a linear composition profile $\overline S_0 (z) = S_0 + \delta S_0(1 - z/H)$, with $\delta S_0$ defined such that the solute concentration is larger by a factor of  two at the bottom of the domain. Afterwards, the fluid is destabilized by a constant heat flux $F_0$ at the top boundary that drives the evolution of the system in time.
We choose the magnitude of $F_0$ in terms of the diffusive heat flux that would be present in the fluid if it was just marginally stable against convection
\begin{equation}
F_{\mathrm{crit}} = k \dfrac{\beta}{\alpha} \left|\dfrac{d\overline S_0}{dz}\right| = k \dfrac{\beta}{\alpha} \left(\dfrac{\delta S_0}{H}\right)\, , \label{eq_f_crit}
\end{equation}
i.e., we set $F_0 = f \times F_{\mathrm{crit}}$, where $f$ is a positive number (5.4 and 10.8 in our numerical experiments). The parameter values used in the simulations were chosen to reproduce the experiments of \citet{1964PNAS...52...49T} and are shown in Table \ref{tab_parameters}. Note that the solute diffusivity $\kappa_S$ was increased by an order of magnitude such that $\tau = 0.1$, and the kinematic viscosity $\nu$ was varied to have a set of simulations that covers $Pr=0.1$, $1$, and $7$.

\begin{table}
\centering
\caption{Parameters used in the simulations.} \label{tab_parameters}
\begin{tabular}{@{}lll@{}}
\hline
& Parameter & Value\\
\hline
$H$ & Height ($\rm m$) & 0.25  \\
$L$ & Width ($\rm m$) & 0.25 \\
$\nu$ & Kinematic viscosity  ($10^{-7}\rm m^2\, s^{-1}$) & $0.142$,\, $1.42$,\, $10$  \\
$\kappa_T$ & Thermal diffusivity ($10^{-7}\rm m^2\, s^{-1}$) & $1.42$ \\
$\kappa_S$ & Solute diffusivity ($10^{-7}\rm m^2\, s^{-1}$) & $0.142$ \\
$k$ & Thermal conductivity ($\rm W\, m^{-1}\, K^{-1}$) & 0.6   \\
$\rho_0$ & Background density ($\rm kg\, m^{-3}$) & 1025   \\
$c_P$ & Specific heat capacity ($\rm J\, K^{-1}\, kg^{-1}$) & 4182 \\
$\alpha$ & Thermal expansion coefficient ($\rm K^{-1}$)  & $2.3\times 10^{-4}$ \\
$\beta$ & Solute contraction coefficient (1)  & $7.6\times 10^{-4}$ \\
$T_0$ & Background temperature ($\rm K$)  & 293.15   \\
$S_0$ & Background solute ($\rm g\, kg^{-1}$)  & 12.78 \\
$\delta S_0$ & Initial solute contrast across depth  ($\rm g\, kg^{-1}$)  & 13  \\
$F_{\rm crit}$ & Critical heat flux for stability ($\rm W\, m^{-2}$)  & 103  \\
$F_0$ & Heat flux at the top boundary ($\rm W\, m^{-2}$)  & 5.4$F_{\rm crit}$, 10.8$F_{\rm crit}$ \\
\hline
\end{tabular}
\end{table}

Since we are interested in the early evolution of the system, our numerical experiments were performed until $t \approx 4500$ s (i.e., $t =  0.01\, t_{\rm diff}$, where $t_{\rm diff}$ is the thermal diffusion time across the box). This is enough time to observe the formation of the outer convective layer and its inwards propagation before the formation of secondary layers.

We solve linear terms implicitly and nonlinear terms explicitly using an implicit-explicit (IMEX), third-order, four-stage Runge-Kutta time-stepping scheme RK443 with the Dedalus spectral code \citep{2020PhRvR...2b3068B}. The variables are decomposed on a Chebyshev (vertical) and Fourier (horizontally-periodic) domain in which the physical grid dimensions are 3/2 the number of modes. Based on a resolution study, we find that 512 modes in each direction is enough to resolve all the fluid flows given the parameters used in this work. However, for a better resolution of small scale structures, we use 1024 modes in each direction.

Although we solve the equations in dimensional form, most of the relevant parameters analysed and presented in this work are dimensionless. Further, for a better interpretation of the results, when plotting the quantities that are not dimensionless, we show them normalized to relevant reference values. For example, the thickness of the convective layer is presented in terms of the height of the box ($H$), time is presented in terms of the thermal diffusion time across the box ($t_{\rm diff} = H^2/\kappa_T$), and temperature and solute are presented in terms of the initial temperature and initial solute contrast across the box ($T_0$ and $\delta S_0$, respectively). Further, the heat fluxes are presented in terms of $F_0$, and solute fluxes in terms of the initial solute flux across the box ($\rho_0 \kappa_S |d\overline S_0/dz|$).  For the interested reader, we present in Sect. \ref{dimensionless} a set of dimensionless equations with the relevant dimensionless parameters that control our simulations.

\subsection{Dimensionless Parameters} \label{dimensionless}

In the following, we non-dimensionalize the Boussinesq equations presented above such that length is in units of the box height ($H$), time is in units of the thermal diffusion time across the box ($H^2/\kappa_T$), solute is units of the initial solute contrast across the box $(\delta S_0)$, and temperature is in units of the imposed flux as $F_0 H/k$. By these choices, velocity is in units of $\kappa_T/H$, and pressure has units of $\rho_0 \kappa_T^2/H^2$. The resulting dimensionless equations are

\begin{gather}
\nabla \cdot \bm{\tilde v} = 0\, ,\\
\dfrac{\partial \tilde T}{\partial \tilde t}  = - (\bm{\tilde v}\cdot \nabla)\,\tilde T + \nabla^2 \tilde T\, ,\\
\dfrac{\partial \tilde S}{\partial \tilde t} = - (\bm{\tilde v}\cdot \nabla)\,\tilde S + \tau\nabla^2 \tilde S\, ,\\
\dfrac{\partial \bm{\tilde v}}{\partial \tilde t} = - (\bm{\tilde v}\cdot \nabla)\,\bm{\tilde v} - \nabla \tilde P + \mathcal{R}_T Pr \left[ \tilde{T} -
\left(\frac{F_0}{F_\mathrm{crit}}\right)^{-1} \tilde{S} \right]\bm{\hat z} + Pr\nabla^2 \bm{\tilde v}\,,
\end{gather}
with boundary conditions
\begin{gather}
\tilde w\,\big\vert_{\tilde z=0,1} = 0\,, \hspace{0.2cm} \dfrac{\partial  \tilde u}{\partial \tilde z}\,\bigg\vert_{\tilde z=0,1} = 0\,, \hspace{0.2cm} \dfrac{\partial \tilde S}{\partial \tilde z}\,\bigg\vert_{\tilde z=0,1} = 0\,,\\
\dfrac{\partial \tilde T}{\partial \tilde z}\,\bigg\vert_{\tilde z=0} = 0\, , \hspace{0.2cm} \dfrac{\partial \tilde T}{\partial \tilde z}\,\bigg\vert_{\tilde z=1} = -1\, .
\end{gather}

We clarify that dimensionless variables are written with a tilde and they should not be confused with horizontally-averaged ($x$-independent) variables, which are written with a line on the top.

The dimensionless parameters that control the simulations are $F_0/F_{\rm crit}$, the Prandtl number ($Pr$), the diffusivity ratio ($\tau$) and a modified Rayleigh number ($\mathcal{R}_T$), defined respectively as
\begin{align}
&\dfrac{F_0}{F_{\rm crit}} = F_0\left(k\dfrac{\beta}{\alpha}\dfrac{\delta S_0}{H}\right)^{-1}\, ,\\
&Pr = \dfrac{\nu}{\kappa_T}\, ,\\
&\tau = \dfrac{\kappa_S}{\kappa_T}\, ,\\
&\mathcal{R}_T = \dfrac{\alpha g H^3}{\kappa_T \nu }\left(\dfrac{F_0 H}{k}\right).
\end{align}

Note that $\mathcal{R}_T (F_0/F_{\rm crit})^{-1}$ can be re-written as

\begin{align}
\mathcal{R}_T \left(\dfrac{F_0}{F_{\rm crit}}\right)^{-1} = \mathcal{R}_S = \dfrac{\beta g H^3 \delta S_0}{\kappa_T \nu },
\end{align}
which looks as the traditional Rayleigh-number but for solute. In terms of the non-dimensionalization described above, the parameters used in our set of six simulations are given in Table \ref{Table_a1} 

\begin{table}
\centering
\caption{Dimensionless parameters used in the simulations.} \label{Table_a1}
\begin{tabular}{@{}lllllll@{}}
\hline
$\#$ & $\tau$ & $Pr$ & $F_0/F_{\rm crit}$ & $\mathcal{R}_T$ & $\mathcal{R}_S$ & $\nu$ ($\rm{10^{-7} m^{2}\, s^{-1}}$)\\
\hline
1 & 0.1 & 0.1 & 5.4 & $4\times 10^{12}$ & $7.5\times 10^{11}$ & $0.142$ \\ 
2 & 0.1 & 0.1 & 10.8 & $8\times 10^{12}$ &  $7.5\times 10^{11}$ & $0.142$ \\ 
3 & 0.1 & 1 & 5.4 & $4\times 10^{11}$ &  $7.5\times 10^{10}$ & $1.42$ \\ 
4 & 0.1 & 1 & 10.8 & $8\times 10^{11}$ &  $7.5\times 10^{10}$ & $1.42$\\ 
5 & 0.1 & 7 & 5.4 & $5.76\times 10^{10}$ & $1.06 \times 10^{10}$ & $ 10$\\ 
6 & 0.1 & 7 & 10.8 & $1.15\times 10^{11}$ & $1.06 \times 10^{10}$ & $10$\\ 
\hline
\end{tabular}
\end{table}

It is worth mentioning that in this problem convection is driven by the temperature difference across the thermal boundary layer due to the imposed heat flux at the top, and the convective layer grows in time. This means that within the convection zone, the classic Rayleigh number and Reynolds number have a time-dependent magnitude determined by the thickness of the convective layer

\begin{equation}
Ra = \dfrac{\alpha g h^3 \delta T}{\kappa_T \nu}\, , \hspace{0.5cm} Re = \dfrac{v_c h }{\nu}\,,
\end{equation}
where $v_c$ is the convective velocity. By measuring $\delta T$, $h$, and $v_c$ at each time directly from the simulations, we find $Ra$ and $Re$ varying from 0 (initially) until a maximum value of $10^9$, and $10^5$, respectively.

\section{Inwards propagation of the convective layer} \label{sec_inwards_prop}
\begin{figure*}
\centering
\includegraphics[width=8.5cm]{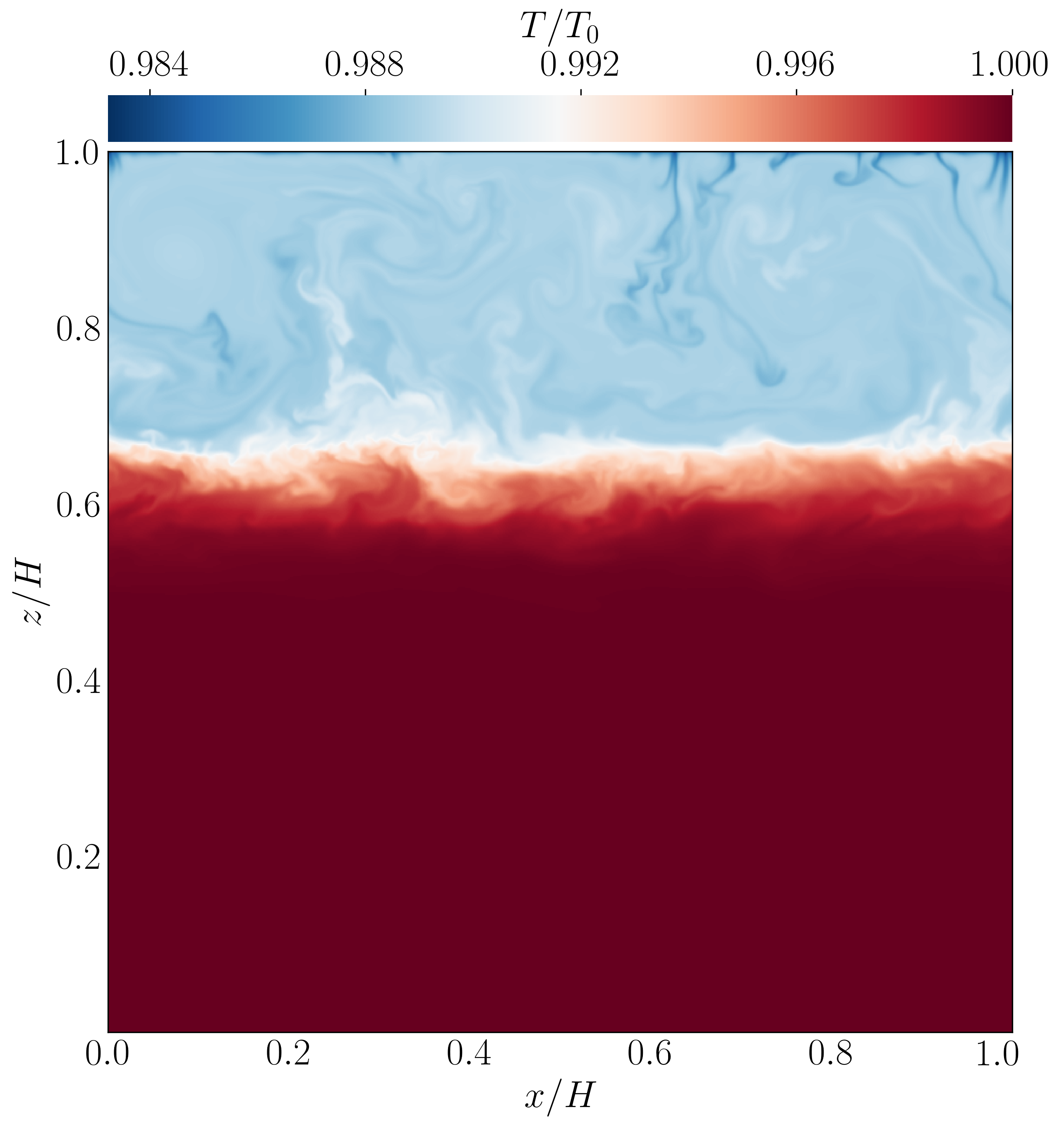}\hspace{0.5mm}
\includegraphics[width=8.5cm]{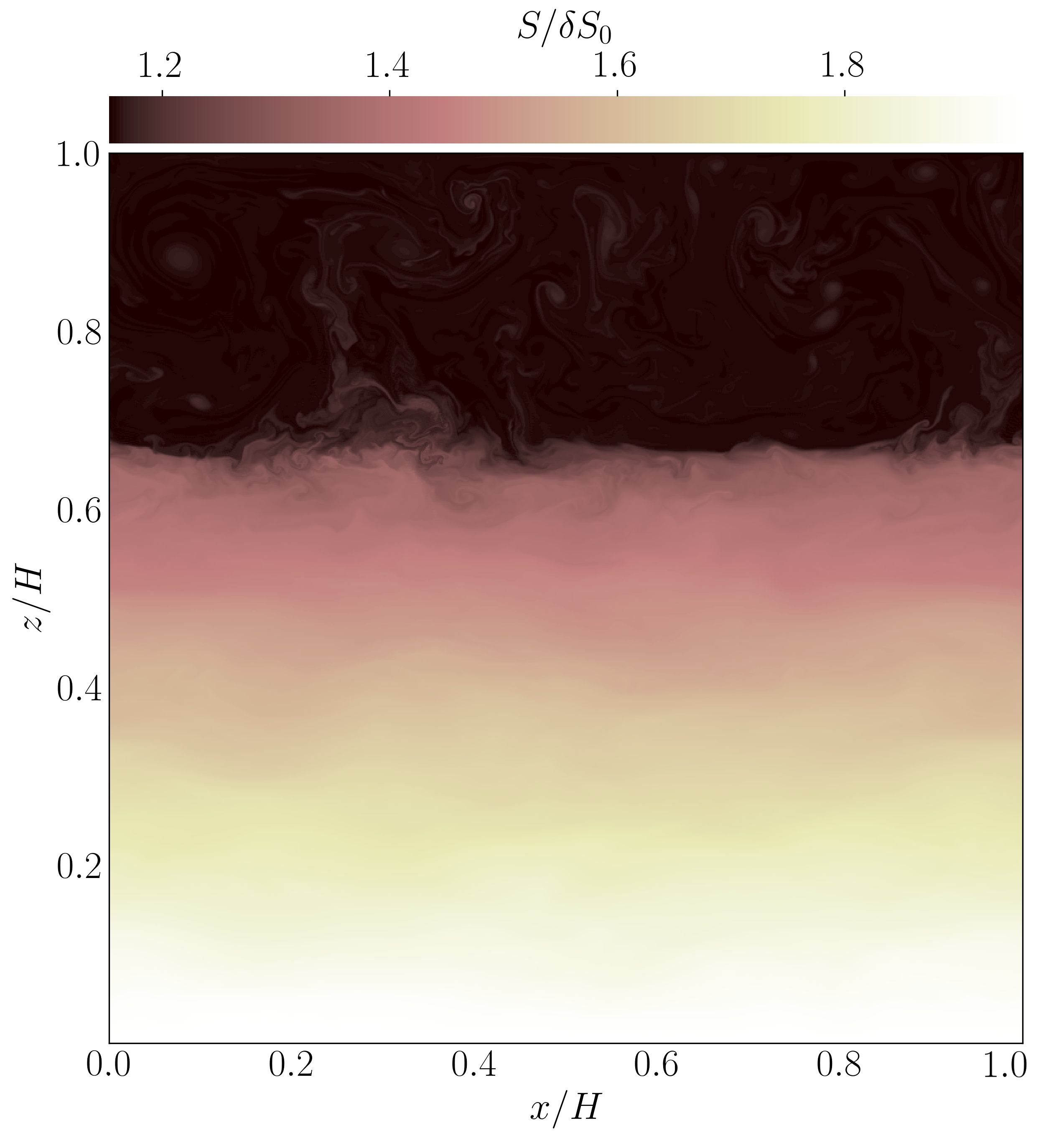}
\caption{Instantaneous snapshots of the temperature field (divided by the initial temperature $T_0$, left panel) and solute field (divided by the initial solute contrast $\delta S_0$, right panel) for the case $Pr=0.1$ and $F_0 = 5.4 F_{\rm crit}$, at $t = 0.005\, t_{\rm diff}$. Blue (red) color represents low (high) temperature. Dark (light) color represents low (high) solute concentration. Note how convective eddies impinging on the interface incorporate fluid from below.} \label{fig_solute_field}
\end{figure*}

We find that the initial behaviour of the system is qualitatively similar for all the simulations: after turning on the heat flux at the top, the cooling rate is high enough that a convective layer, well mixed in both temperature and composition, quickly forms and grows inwards by incorporating fluid from below, as shown in the snapshots in Fig.~\ref{fig_solute_field}.

\begin{figure*}
\centering
\includegraphics[width=17cm]{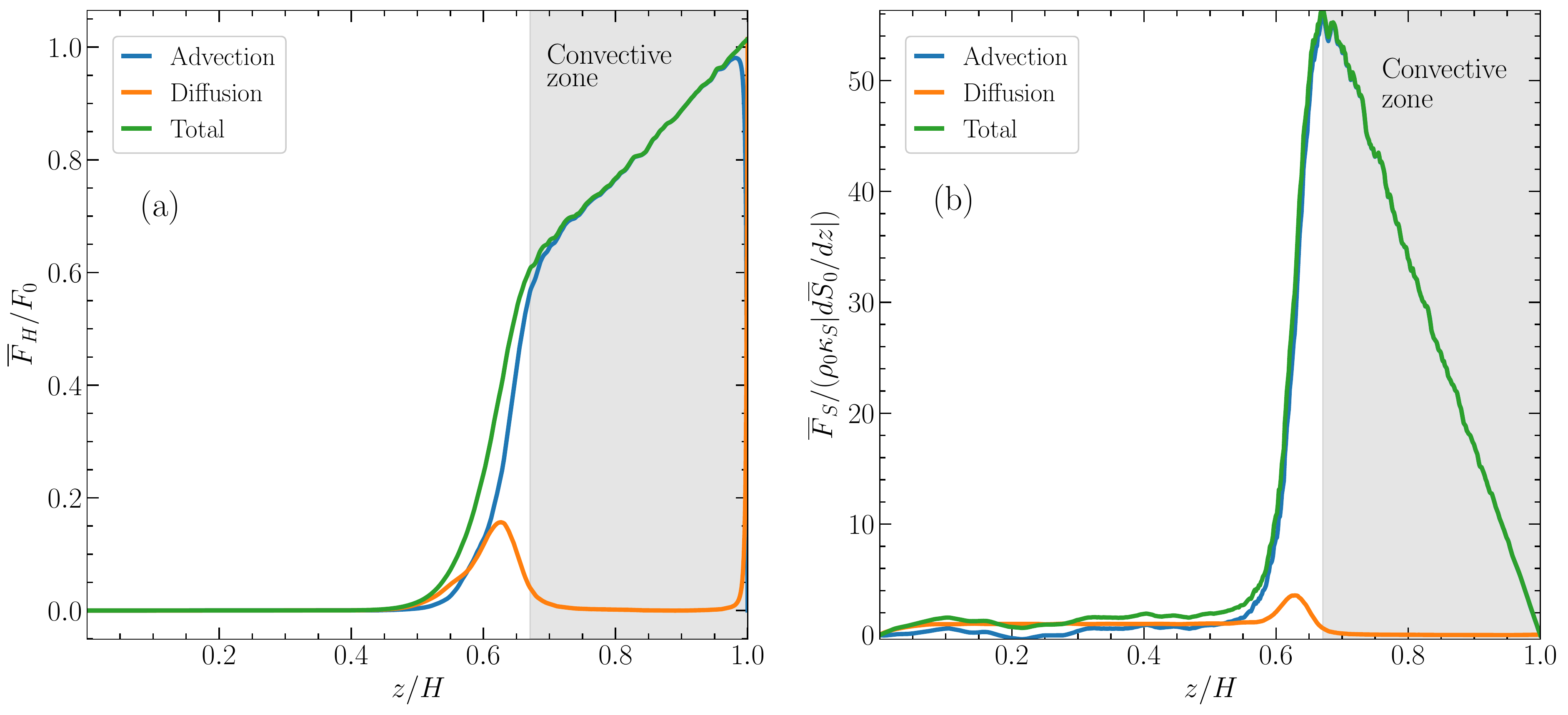}
\caption{Horizontally-averaged flux profiles for the simulation $Pr=0.1$ and $F_0 = 5.4 F_{\rm crit}$ at $t=2280$ s ($t=0.005\, t_{\rm diff}$). Panels (a) and (b) show profiles of heat and solute flux, respectively. In both panels, the green, blue, and orange lines correspond to the total, advective, and diffusive contribution to the flux, respectively. The shaded areas denote the extent of the convective zone. Note that in panel (a) the heat fluxes are normalized to $F_0$, and in panel (b) the solute fluxes are normalized to $\rho_0 \kappa_S |d\overline S_0/dz|$. Further, in both panels the $z$ coordinate is normalized to the height of the box ($H$).} \label{fig_flux_profiles}
\end{figure*}

To get some intuition on how temperature and composition change within the convective layer, we look into the horizontally-averaged profiles of heat and solute fluxes, which we define as
\begin{align}
&\overline{F}_H = \rho_0 c_P \overline{wT} - k\, d\overline T/dz\,  \label{eq_heat_flux}\\ 
&\overline{F}_S = \rho_0\overline{wS} - \rho_0\kappa_S\, d\overline S/dz\, , \label{eq_salinity_flux}
\end{align}
respectively. The first and second term on the right hand side in Eqs.~\eqref{eq_heat_flux} and \eqref{eq_salinity_flux} correspond to the advective and diffusive fluxes, respectively.  As an example, we show in Fig.~\ref{fig_flux_profiles} the flux profiles for the case $Pr=0.1$ and $F_0 = 5.4 F_{\rm crit}$ at $t=2280$ s ($t=0.005\, t_{\rm diff}$), the same snapshot as shown in Fig.~\ref{fig_solute_field}. Despite the fluctuations due to the advective contribution to the fluxes, it is clear that in the convective layer the total heat flux increases linearly with depth (Fig.~\ref{fig_flux_profiles}a), meaning that the fluid is cooling everywhere at a constant rate to keep its temperature uniform. A similar behaviour is observed in the composition flux (Fig.~\ref{fig_flux_profiles}b). In the convective zone the total flux decreases linearly with depth, thereby, the solute content is increasing everywhere at the same rate to keep the fluid with uniform composition.

\begin{figure}
\centering
\includegraphics[width=8cm]{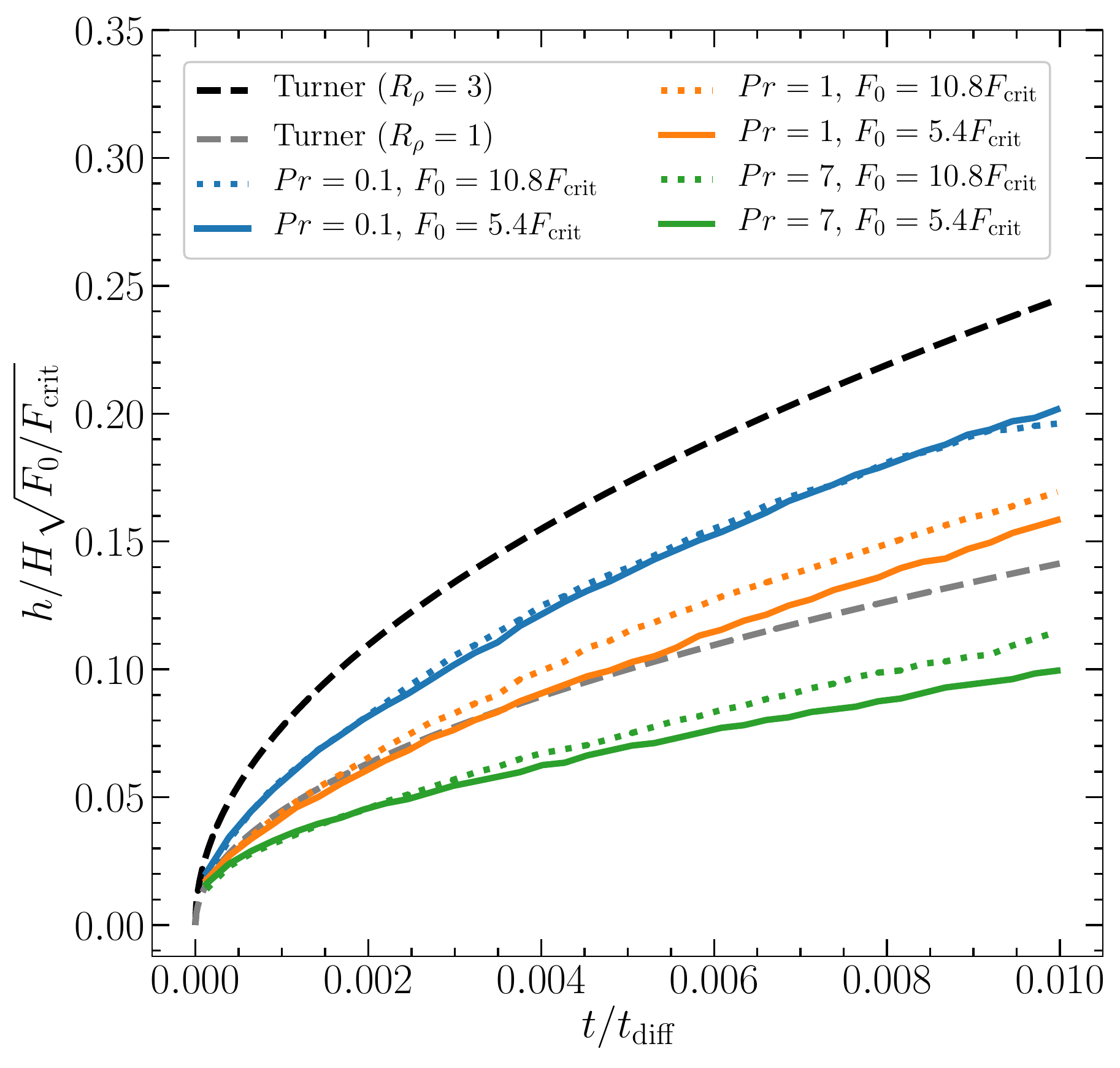}
\caption{Temporal evolution of the thickness of the outer convective layer, $h$, divided by $H\sqrt{F_0/F_{\rm crit}}$. The gray and black dashed-lines correspond to the predictions by Eq. \eqref{eq_h_general} using $R_\rho=1$ and $3$, respectively. Colors distinguish between different $Pr$, and line-style distinguish between different $F_0$ (dotted-lines in the case $F_0 = 10.8 F_{\rm crit}$, and solid-lines for $F_0 = 5.4 F_{\rm crit}$.} \label{fig_thickness}
\end{figure}

Figure \ref{fig_thickness} shows the evolution in time of the thickness of the convective zone. To help compare the different simulations, we remove the $h\propto \sqrt{F_0/F_{\rm crit}}$ scaling predicted by Turner's analytic model (Eq.~\ref{eq_h_general}) by plotting $h/(H\sqrt{F_0/F_{\rm crit}})$. For comparison, we show $h/(H\sqrt{F_0/F_{\rm crit}})$ as predicted by Eq.~\eqref{eq_h_general} using $R_\rho=1$ and $R_\rho=3$.
Comparing the different curves, we see that there is a weak dependence of the rate of growth of the convection zone on $Pr$, such that the convective layer grows faster as $Pr$ decreases. For example, at $t=0.01\, t_{\rm diff}$, the height of the convective zone at $Pr=0.1$ is larger than for $Pr=7$ by a factor of two. Comparing curves at the same $Pr$, we see also that the growth rate of the convective layer increases slightly faster with flux than the expected $\sqrt{F_0/F_{\rm crit}}$ scaling. This can be seen in Fig.~\ref{fig_thickness} where the curves for $F_0= 10.8 F_{\rm crit}$ lie slightly above those for $F_0=5.4F_{\rm crit}$. The maximum deviations between the curves for different fluxes are $2.8\%, \, 6.4\%$, and $13.2\%$, for $Pr = 0.1,\, 1$ and $7$, respectively. As we discuss below, the variations with $Pr$ can be understood in terms of differences in the entrainment efficiency with $Pr$, as well as the effect of the heat flux at the boundary between the convection zone and stable layer, which is not included when deriving Eq.~\eqref{eq_h_general}.

The best-fit power law to the convection zone depth as function of time is close to but not exactly $h\propto t^{1/2}$. Fitting a general power law to the data, we find $h\propto t^{0.467(5) - 0.585(2)}$, where the lowest and highest rate correspond to the cases ($Pr=7$, $F_0 = 5.4F_{\rm crit}$), and ($Pr=0.1$, $F_0 = 10.8F_{\rm crit}$), respectively (the values in parenthesis correspond to the uncertainties in the last digit). For $Pr=7$,  \cite{fernando_1987} and \cite{molemaker_dijkstra_1997} found that their data was fit by $h \propto t^{0.36 - 0.5}$ depending on the magnitude of the imposed flux $F_0$.

\begin{figure*}
\includegraphics[width=17cm]{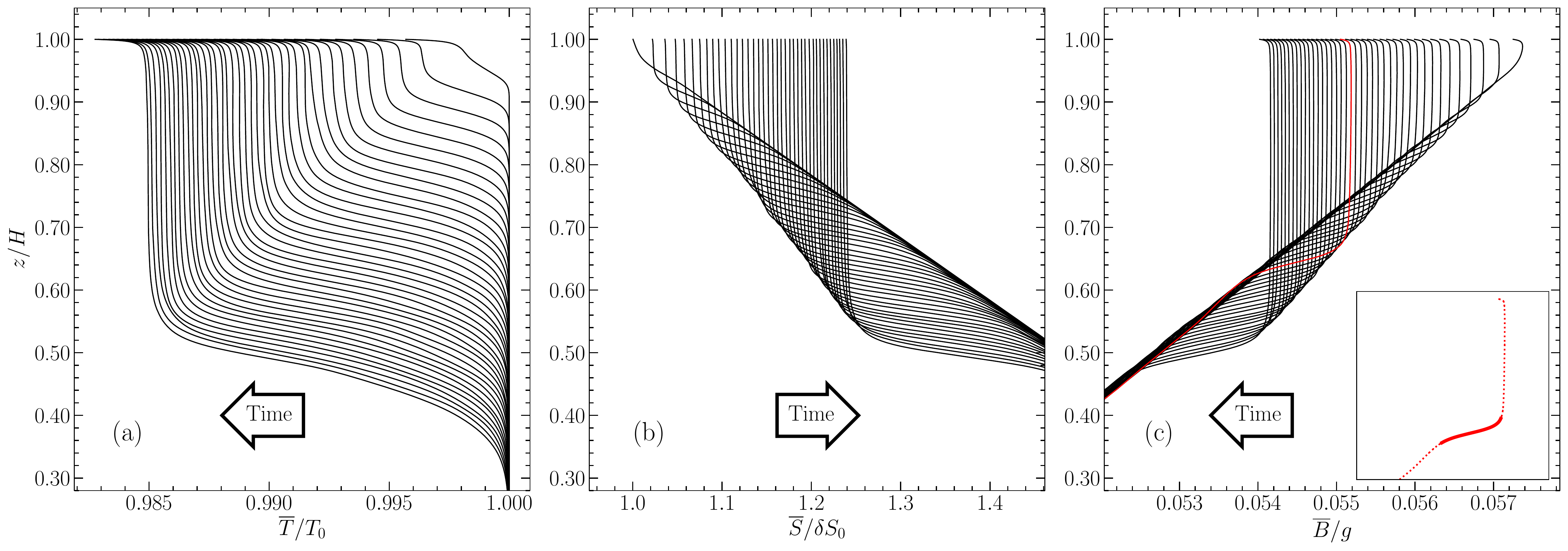}
\caption{Horizontally-averaged profiles of the temperature (normalized to $T_0$), solute (normalized to $\delta S_0$), and buoyancy field (normalized to the magnitude of the acceleration due to gravity, $g$), in panels a, b, and c, respectively. The results corresponds to the simulation at $Pr = 0.1$ and $F_0 = 5.4 F_{\rm crit}$. Profiles are shown at different times (where time evolves according the direction of the arrows). All panels share the same scale along the $z$-axis. Note that the $z$ coordinates is hormalized to the height of the box, $H$. In panel (c), the buoyancy profile at $t = 0.005\, t_{\rm diff}$ is shown in red and a zoomed-region shows the buoyancy step using a thicker line.} \label{fig_profiles}
\end{figure*}

\section{Entrainment at the convective boundary} \label{sec_entrainment}

In this section, we investigate entrainment at the convective boundary as the mechanism responsible for mixing and growth of the convective layer at $Pr \leq 1$. In particular, we show that: 1) during the propagation of the convective layer, a buoyancy jump across the interface is present, which suggests that a process is needed to transport heavier fluid across the stable interface; 2) the entrainment equation proposed and tested by \citet{fernando_1987} and \citet{molemaker_dijkstra_1997} in experiments and simulations of salty water ($Pr=7$) gives a good description of our results at lower $Pr$.

\subsection{Development of a buoyancy jump in a stable interface}

Figure \ref{fig_profiles} shows horizontally-averaged profiles of the temperature, $\overline T$, composition, $\overline S$, and buoyancy, $\overline B = g(\alpha \overline T - \beta \overline S)$, at different times for the case $Pr = 0.1$ and $F_0 = 5.4 F_{\rm crit}$. A buoyancy jump at the base of the convection zone develops and persists over time. We found the same behavior in all our simulations. To show this more clearly, we show in the inset of panel (c), the profile at a particular time, with the region denoting the buoyancy jump using a thicker red line.

\begin{figure*}
\centering
\includegraphics[width=14cm]{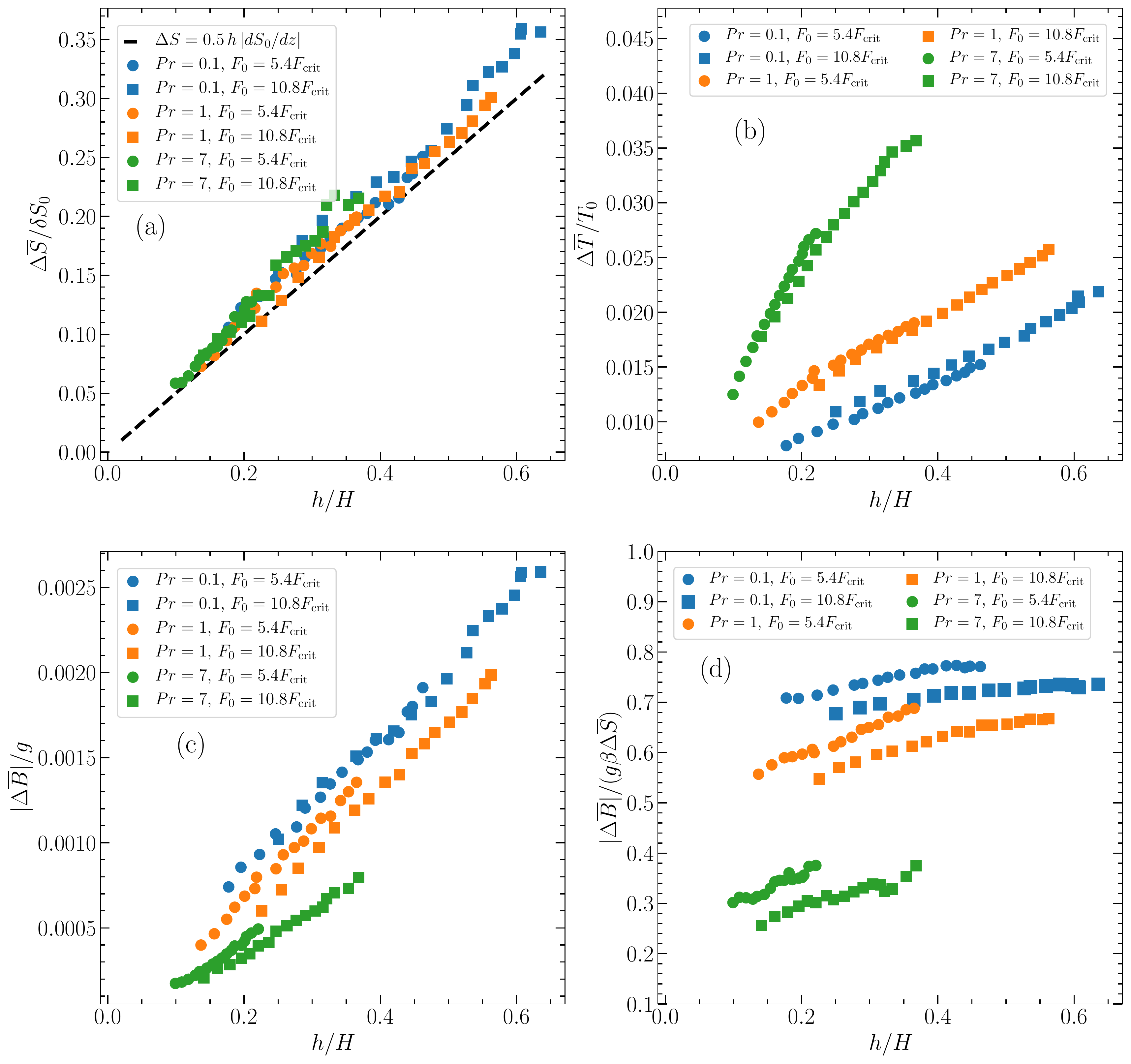}
\caption{Jumps at the interface as a function of $h/H$ (the thickness of the convective layer normalized to the height of the box). Panels (a), (b), and (c) show the absolute jumps of composition (normalized to $\delta S_0$), temperature (normalized to $T_0)$, and buoyancy (normalized to the magnitude of the acceleration due to gravity, $g$), respectively. Panel (d) shows the ratio $|\Delta \overline B|/g\beta \Delta \overline S$ versus $h/H$.  As shown in the legends, colors distinguish between different $Pr$, and markers distinguish between different values of $F_0$}. \label{fig_jumps}
\end{figure*}

Figure ~\ref{fig_jumps} shows the jumps in solute, temperature, and buoyancy across the interface ($\Delta \overline S$, $\Delta \overline T$, and $\Delta \overline B$, respectively) as a function of the thickness of the convective layer, $h$. We measure the jump in each quantity from horizontally-averaged profiles, defined as the value below the interface (stable region) minus the value above the interface (convective region), so that $\Delta \overline T$ and $\Delta \overline S$ are positive quantities, whereas $\Delta \overline B$ is negative for a stable interface. It is worth mentioning that the dispersion in our measurements is due to the propagation of waves near the interface, which make its location (start and end) time-variable, especially in the simulated experiment with $Pr=0.1$ and $F_0 = 10.8 F_{\rm crit}$. 

We observe that the jumps in solute, temperature and buoyancy all exhibit a monotonic (positive) trend with $h$, weakly dependent on $F_0$. As expected, since solute is conserved during the evolution of the convective layer, $\Delta \overline S$ exhibits a linear trend with $h$ (Eq. \ref{eq_salinity_conservation}), independent of $Pr$ and $F_0$. The situation for $\Delta \overline T$ is less clear and there are substantial differences between the simulations, probably due to the effect of heat flux at the interface between the convective layer and stable region. The buoyancy jump $\Delta \overline B$ also exhibits a linear trend with $h$, but its magnitude is larger for simulations at $Pr=0.1$. It is interesting that the ratio $|\Delta \overline B|/g\beta \Delta \overline S$ increases slowly with $h$, being roughly constant for each experiment. We clarify that roughly constant means maximum variations at the level of $20\%$. We find that the solute difference across the interface accounts for 20-80$\%$ of the buoyancy jump, depending on $Pr$ and $F_0$. For comparison, \citet{fernando_1987} and \citet{molemaker_dijkstra_1997} found for salty water that the salinity jump across the interface accounts for $11\%$ and $50\%$ of the buoyancy jump, respectively. The differences can be explained by the magnitude of the imposed heat flux and the initial solute gradient. In terms of our units, \citet{fernando_1987} and \citet{molemaker_dijkstra_1997} used $F_0 \approx 18 F_{\rm crit} $ and $F_0 \approx 5.6 F_{\rm crit}$, respectively.

\subsection{Entrainment equation and mixing efficiency} \label{sec_entraiment_mixing}

A key parameter of the entrainment mechanism is the so-called mixing efficiency. The entrainment hypothesis states that the rate of change of potential energy due to mixing is proportional to the kinetic energy flux available near the interface \citep[e.g.,][]{linden_1975}. By assuming that a constant fraction of the available kinetic energy is used to lift heavier fluid across the interface, \citet{fernando_1987} and \citet{molemaker_dijkstra_1997} derived an expression for the rate of change of the convection zone thickness 
\begin{equation}
\dfrac{(-\Delta \overline B)}{g} \dfrac{dh}{dt} = \gamma \left(\dfrac{\alpha F_{0}}{\rho_0 c_P}\right)\, ,\label{eq_entrainment_balance}
\end{equation}
which defines the mixing efficiency, $\gamma$ \citep[see, e.g., the discussion in Sect. 3.2.4 in][]{molemaker_dijkstra_1997}. The entrainment rate is often also written in terms of a bulk Richardson number $Ri=h\Delta \overline B/v_c^2$, where $v_c$ is the rms convective velocity and we use the height of the convective layer as the length-scale of the turbulent motions. Using mixing-length theory to write $v_c \sim (g \alpha F_0/\rho_0 c_P)^{1/3} h^{1/3}$, Eq.~(\ref{eq_entrainment_balance}) takes the form $dh/dt\approx \gamma v_c/Ri$. 
We find $10 \lesssim Ri\ \lesssim 100$ for all our numerical experiments. Our results fall within the same parameter range reported in previous laboratory experiments of turbulent entrainment \citep
{fernando1991,1997JFM...347..235M}, and hydrodynamics simulations of stellar convective boundaries \citep{Meakin2007}. This corresponds to the intermediate stability regime in which the convective zone expands and the interface is moderately distorted by convective eddies. For much larger values of $Ri$ the entrainment process weakens and the evolution of the interface is expected to be controlled by diffusive processes \citep{fernando_1987,molemaker_dijkstra_1997,2001PhDT.........8B}. 

\citet{fernando_1987} and \citet{molemaker_dijkstra_1997} found in their experiments at $Pr = 7$ that $\gamma$ increases slowly with time, with maximum variations at the level of $30\%$. They reported time-averaged values of $\gamma$ between 0.15-0.56 depending on the magnitude of the imposed heat flux at the boundary. In the following, we test whether $\gamma$ exhibits a similar behaviour at lower $Pr$.

\begin{figure}
\centering
\includegraphics[width=8cm]{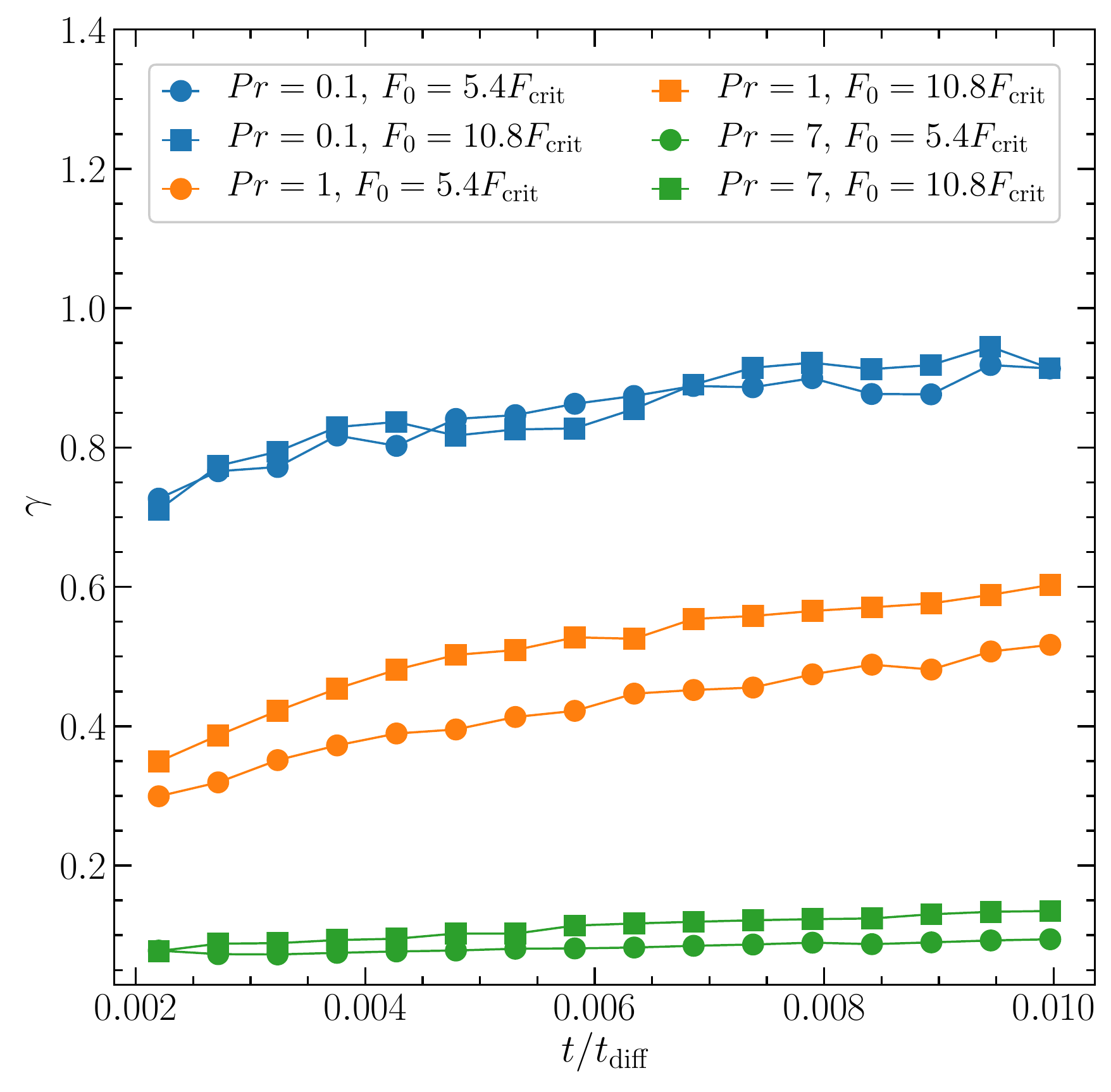}
\caption{Entrainment parameter $\gamma$ as a function of $t/t_{\rm diff}$ (i.e., time normalized to the thermal diffusion time across the box) for all our numerical simulations. As shown in the legends, colors distinguish between different $Pr$, and markers distinguish between different $F_0$. For simulations at $F_0 = 5.4 F_{\rm crit}$, the averaged values of $\gamma$ for $Pr = (0.1,\, 1,\, 7$) are $\gamma \approx (0.85,\, 0.44,\, 0.08)$, whereas for simulations at $F_0 = 10.8 F_{\rm crit}$, the averaged values are $\gamma \approx (0.84,\, 0.51,\, 0.12$).} \label{fig_gamma_times}
\end{figure}

We compute $\gamma$ at different times by using the buoyancy jumps $\Delta \overline B$ from horizontally-averaged profiles, as the ones in Fig.~\ref{fig_profiles}, and $dh/dt$ from differentiation of a power law fit to the curves $h(t)$ in Fig.~\ref{fig_thickness}. Despite the dispersion due to measurement uncertainties in $\Delta \overline B$, the evolution of $\gamma$ behaves similarly at low and high $Pr$, increasing slowly with time, with maximum variations at the level of 20-40 $\%$ (Fig.~\ref{fig_gamma_times}). We find that the time-averaged values of $\gamma$ take values between $0.08$ and $1$, being higher at low $Pr$ and high $F_0$. The trend with $F_0$ is less clear at $Pr=0.1$ since the flow is more turbulent and the dispersion in the measurements is higher.  Our results make sense given that a larger value of $F_0$ provides more energy to the convective eddies, thereby they can entrain and mix more efficiently. Furthermore, low $Pr$ fluids have strong velocity gradients near the interface, enhancing shear motions and mixing. Finally, low $Pr$ fluids are more turbulent (have a larger Reynolds number) and deliver energy to smaller scales with the result that entrainment might be expected to be more efficient. Note that decreasing $Pr$ at a fixed thermal diffusivity means that the thickness of the viscous boundary layer that separates the convective layer and the static fluid below gets smaller, thereby convective eddies entrain through a thinner layer, mixing the fluid more easily.

Our results compare reasonably well with previous work. Our measurements of $\gamma$ for simulations at $Pr=7$ ($\gamma \approx 0.08$ for $F_0 = 5.4 F_{\rm crit}$, and $\gamma \approx 0.12$ for $F_0 = 10.8 F_{\rm crit}$) are expected to be smaller than those reported in \citet{fernando_1987}, who obtained $\gamma \approx 0.5$ for $F_0 \approx 18 F_{\rm crit}$. However, for the case $F_0 = 5.4 F_{\rm crit}$, we expected consistency with \citet{molemaker_dijkstra_1997}, who obtained $\gamma \approx 0.15$ for $F_0 \approx 5.6 F_{\rm crit}$ in \citet{molemaker_dijkstra_1997}, but our measurement is roughly smaller by a factor of 2.

At this point, we have shown that during the propagation of the convective layer, a buoyancy jump develops over the interface. Further, using the entrainment equation (Eq. \ref{eq_entrainment_balance}), we have shown that $\gamma$ behaves in a similar way at low and high $Pr$, increasing slowly with time. We have also shown that $\gamma$ is higher at low $Pr$ and high $F_0$, which suggests that entrainment is stronger in the more turbulent and energetic flow.

\section{The effect of the interfacial heat flux} \label{sec_flux_below}

For $Pr=7$, \citet{molemaker_dijkstra_1997} pointed out that there is a significant heat flux across the interface between the convection zone and stable layer below. This has the effect of heating the convective layer from below and thereby reducing the rate at which it penetrates into the stable layer. In this section, we present our measurements of the interfacial heat flux as the convective layer evolves, and test whether it is significant at low $Pr$.

The change in the heat content within the convective layer of thickness $h$ \cite{molemaker_dijkstra_1997} is determined by
\begin{equation}
\rho_0 c_P h \dfrac{d \Delta \overline T}{dt} = F_0 - \overline F_H^{\,\rm i}\text{,} \label{eq_heat_transport}
\end{equation}
where $\overline F_H^{\,\rm i} = \rho_0 c_P \Delta \overline T dh/dt + \overline F_{\rm{a}}$ is the total heat flux through the interface. The term $\rho_0 c_P \Delta \overline T dh/dt$ corresponds to heat flux through the interface that results from a change $dh = \dot h dt$ in the thickness of the convective layer, and $\overline F_{\rm a}$ is additional heat flux from below. Note that with $\overline{F}_{\rm a}=0$, Eq.~(\ref{eq_heat_transport}) reduces to Eq.~(\ref{eq_heat_conservation}). We measure $\overline F_H^{\,\rm i}$ from the flux profiles in Fig \ref{fig_flux_profiles}(a) as the value of the total heat flux at the edge of the convective zone.

\begin{figure*}
\centering
\includegraphics[width=\textwidth]{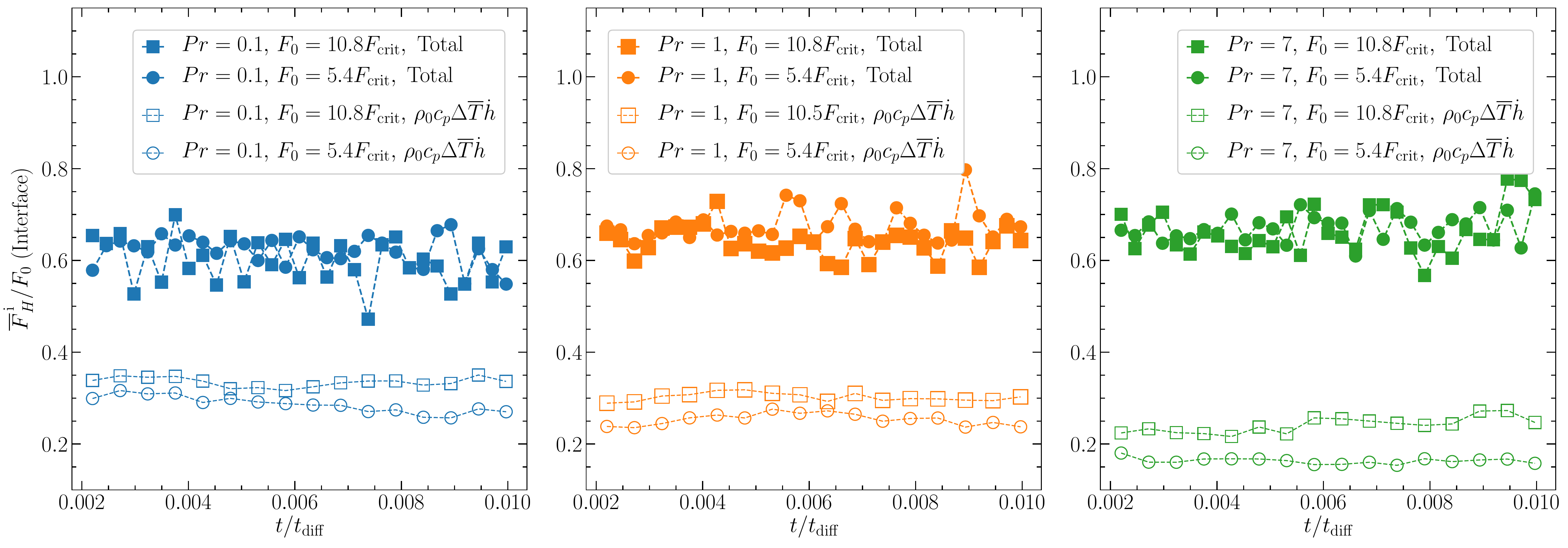}
\caption{Heat flux through the interface (normalized to $F_0$) as a function of $t/t_{\rm diff}$ (time normalized to the thermal diffusion time across the box). As shown in the legends, colors distinguish between different $Pr$, and markers distinguish between different $F_0$. All panels share the same scale in both axes. In all panels, filled and unfilled colors distinguish between the total heat flux and the resulting flux due to a change $dh = \dot h dt$ in the convective zone, respectively}. \label{fig_flux_interface}
\end{figure*}

Figure \ref{fig_flux_interface} shows for all our simulations the temporal evolution of the total heat flux through the interface, $\overline F_H^{\,\rm i}$, normalized to the imposed cooling flux $F_0$. For comparison, we also include the contribution of the $\rho_0 c_P \Delta \overline T dh/dt$ term. Interestingly, we find that $\overline F_H^{\,\rm i}$ is weakly-dependent of $Pr$ and $F_0$, and it fluctuates around a constant value $\approx 0.6 F_0$. The contribution from $\rho_0 c_P \Delta \overline T \dot{h}$ also fluctuates around a constant value but it is slightly different depending on $F_0$ and $Pr$.  We subtract $\rho_0 c_P \Delta \overline T \dot{h}$ from $\overline F_H^{\,\rm i}$, and take the temporal average between $1000 - 4500\ {\rm s}$ to quantify $\overline F_{\rm{a}}$ for all our simulations. 

\begin{figure}
\centering
\includegraphics[width=8cm]{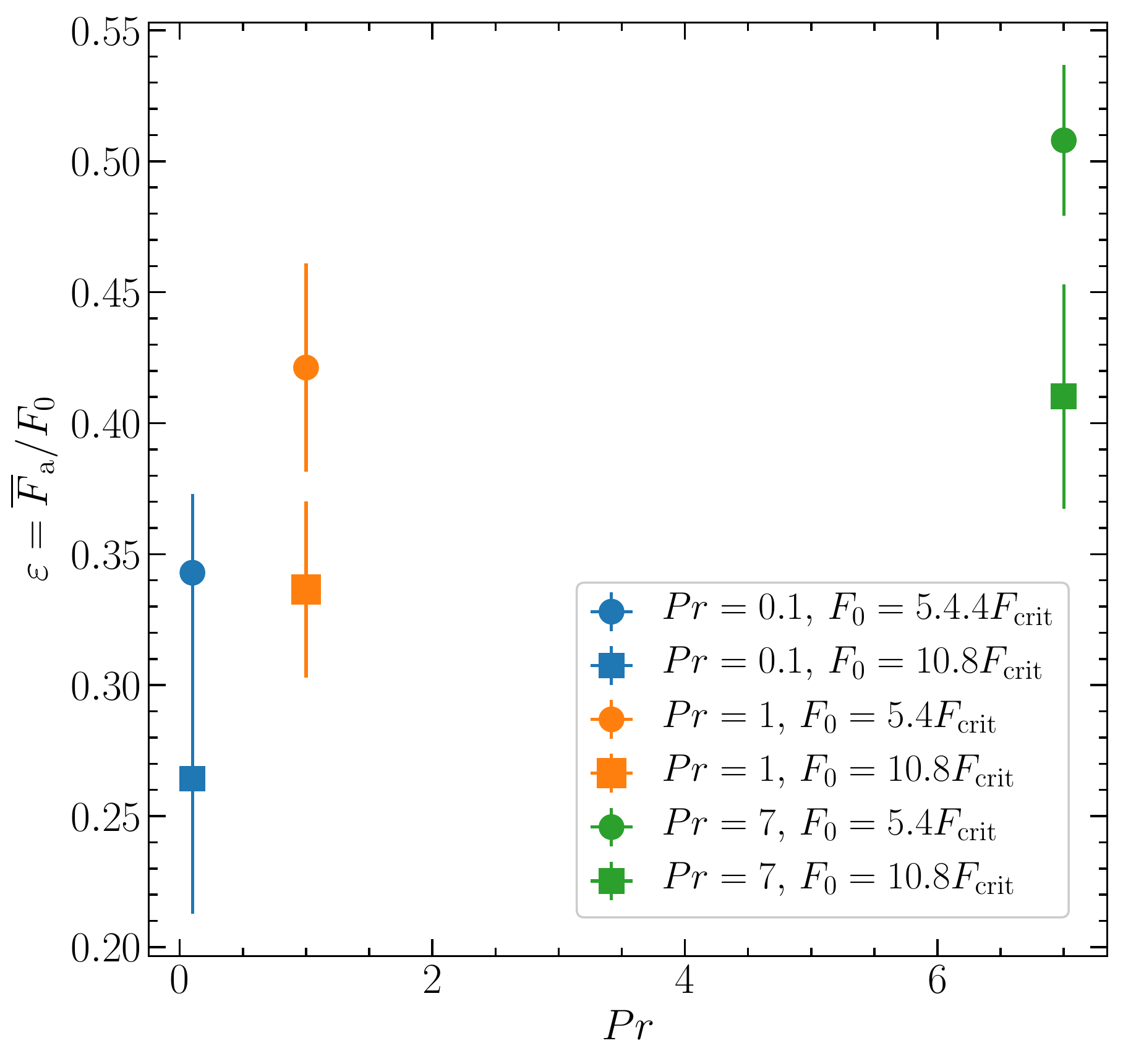}
\caption{Ratio $\overline F_{\rm{a}}/F_0$ as a function of $Pr$. As shown in the legends, colors distinguish between different $Pr$, and markers distinguish between different $F_0$ (for a given $Pr$). The error bars correspond to one sigma from the mean.} \label{fig_f_h_in}
\end{figure}

We find that $\overline F_{\rm{a}}$ is a fixed fraction of the imposed heat flux at the top, $\overline F_{\rm{a}}=\varepsilon F_0$, with $\varepsilon$ varying between $0.25$ and $0.5$, therefore it significantly affects the growth rate of the convection zone (Fig.~\ref{fig_f_h_in}). Further, we observe that $\overline F_{\rm{a}}$ increases with $Pr$ and for all the simulations at $F_0 = 5.4 F_{\rm crit}$,  it is $\approx 25\%$ larger than for $F_0 = 10.8 F_{\rm crit}$. This result makes sense because at high $Pr$ the thickness of the convective layer is smaller, thereby the temperature of the convective layer drops more quickly. This implies a higher temperature contrast with the fluid below (Fig.~\ref{fig_jumps}b), resulting in more diffusion of heat upwards. The fact that increased $\overline F_{\rm{a}}$ slows the convection zone growth is consistent with the curves of $h(t)/H\sqrt{F_0/F_{\rm crit}}$ in Fig.~\ref{fig_thickness}, which show that for $F_0 = 10.8 F_{\rm crit}$ the curves lie above the ones for $F_0 = 5.4 F_{\rm crit}$, and the difference between them increases from low to high $Pr$.

\section{Buoyancy transport across the interface}
\label{sec_flux_transport}

\begin{figure*}
\centering
\includegraphics[width=\textwidth]{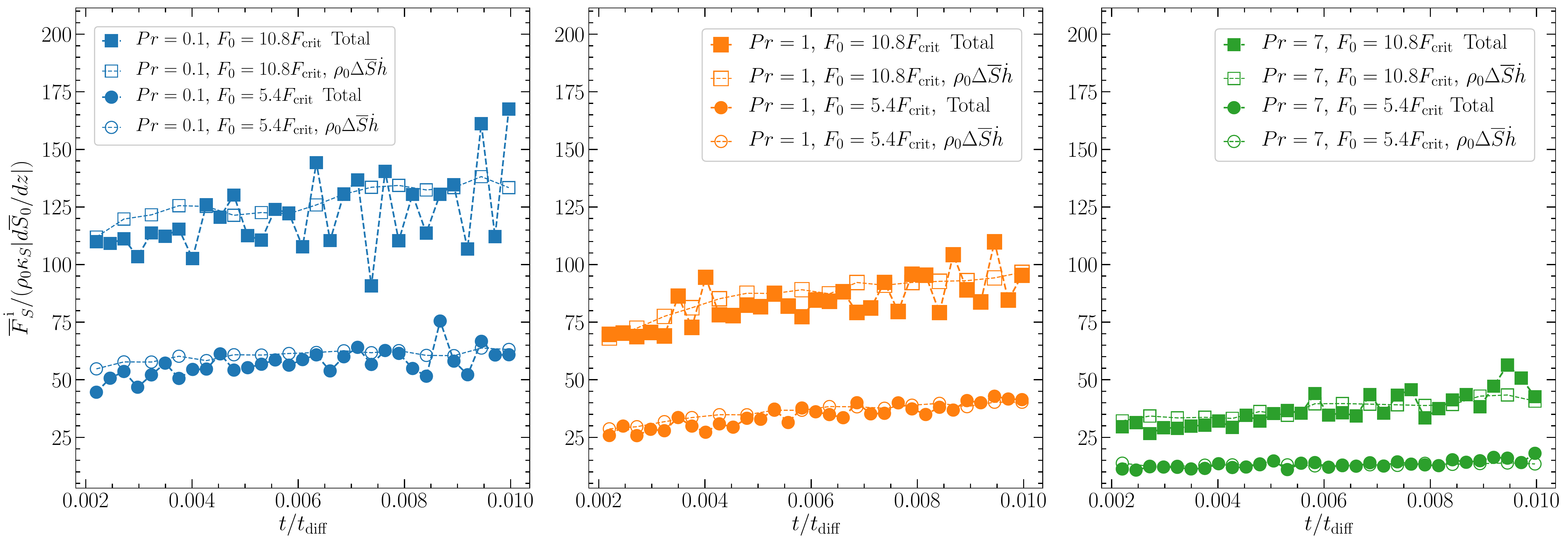}
\caption{Solute flux through the interface (normalized to the initial diffusive solute flux, $\rho_0 \kappa_S |d\overline S_0/dz|$) as a function of $t/t_{\rm diff}$ (i.e., time normalized to the thermal diffusion time across the box). As shown in the legends, colors distinguish between different $Pr$, and markers distinguish between different $F_0$. In all panels, filled and unfilled colors distinguish between the total heat flux and the resulting flux due to a change $dh = \dot h dt$ in the convective zone, respectively.} \label{fig_fluxs_interface}
\end{figure*}

The buoyancy jump at the bottom of the convective layer suggests that there must be a net transport of buoyancy across the interface as the convection zone grows. In this section we investigate the relative heat and solute fluxes at the interface. 
First, similarly to the heat flux in Sect.~\ref{sec_flux_below}, we measured the solute flux at the interface. This is shown in the left panel of Fig.~\ref{fig_fluxs_interface}. We find that the solute flux agrees well with flux implied by the growth rate of the layer, $\rho_0  \Delta \overline S \dot h$. We also observe that the solute transport is higher at low $Pr$ and high $F_0$, consistent with the fact that the convective layer grows faster in these cases. All of these results are consistent with and expected from mass conservation. 

An indication of the nature of the transport at the interface is the relation between the buoyancy flux ratio
\begin{equation}
R_F \equiv \dfrac{\beta \overline F_S^{\,\rm i }}{\alpha \overline F_H^{\,\rm i} c_P^{-1}}\, 
\end{equation}
and the stability of the interface characterized by the density ratio parameter, $R_{\rho} \equiv \beta \Delta \overline S / \alpha \Delta \overline T$ (defined here such that $R_{\rho} > 1$ indicates a stable interface). For example, if the transport were only by diffusion in the interface, the solute and heat fluxes are given by
\begin{align}
\overline F_S^{\,\rm i } \approx \rho_0 \kappa_S \left(\dfrac{\Delta\overline S}{\delta_S}\right)\, ,\hspace{0.25cm} \overline F_H^{\,\rm i } \approx \rho_0 c_P \kappa_T \left(\dfrac{\Delta\overline T}{\delta_T}\right)\, ,
\end{align}
where $\delta_S$ and $\delta_T$ are the thicknesses of the diffusive boundary layers of solute and temperature, respectively. If $\delta_S \approx \delta_T$, this gives
\begin{equation}
R_F = \tau R_\rho\, . \label{eq_ratio_diff_1}
\end{equation}
However, it might be expected that $\delta_S$ and $\delta_T$ would have a different thickness. \citet{fernando_1989} suggested that the interface thickness is set by a balance between the diffusion time across the layer and the convective turnover time. Using mixing length theory for the convective flux and equating it to the diffusive flux across the layer gives
\begin{equation}
R_F = \tau^{1/2} R_{\rho}\,  \label{eq_ratio_diff_2}
\end{equation}
instead.

Experimentally, different relations between $R_F$ and $R_\rho$ have been reported for the transport across a single interface bounded by two convective layers in salty water (rather than an interface between a convective layer and a stable layer as we study here). \citet{turner_1965} found that for $2<R_\rho<7$, the flux ratio $R_F$ is a constant, independent of $R_\rho$. This was confirmed by \citet{linden_shirtcliffe_1978} who found that the value of $R_F$ was consistent with $R_F=(\kappa_S/\kappa_T)^{1/2} = \tau^{1/2}$. Further analysis by \citet{newell_1984} showed that at very large $R_\rho$, the flux ratio obeys Eq.~\eqref{eq_ratio_diff_1}. The differences in the behaviour of $R_F$ were attributed to the nature of the transport across the interface. At low $R_\rho$, advection dominates the fluxes and enhances the transport of salt, whereas at large $R_\rho$, the transport is dominated by molecular diffusion. Note that in the latter case (transport by diffusion), both relations $R_F =\tau
^{1/2} R_\rho$ and $R_F =\tau R_\rho$ have succeeded at explaining different experimental data \citep{newell_1984,turner_1970}. 
More recently, in the context of the transport of heavy elements between the core and the gaseous envelope of Jupiter, \citet{2017ApJ...849...24M} performed three-dimensional simulations for $Pr=\tau=0.03-0.3$. They identified the advective and diffusive regimes of the interface described above, but in both regimes the buoyancy flux ratio was roughly independent of $R_\rho$ and significantly greater than $\tau^{1/2}$.

Fig.~\ref{fig_flux_ratio} shows our measurements of the buoyancy flux ratio as a function of $R_\rho$. 
We find that $R_F$ increases with $R_{\rho}$, so that as the convection zone deepens and the interface becomes more stable (larger $R_\rho$), there is a larger solute flux compared to heat flux.  
As expected, since the total heat flux through the interface is approximately the same for all our experiments, we find that the evolution of $R_F$ scales in the same way as the solute flux $\overline F_S^{\rm i}$, i.e., $R_F$ decreases with increasing $Pr$ and increases with increasing $F_0$. The range of values of $R_F$ seems to converge towards $\tau^{1/2}$ as $Pr$ increases, consistent with the measurements for $Pr=7$ in laboratory experiments. However, in all cases we find that $R_F > \tau^{1/2}$, consistent with the results in \citet{2017ApJ...849...24M}, although our values of $R_F$ are significantly larger than theirs, as shown in Fig.~\ref{fig_flux_ratio}c.
In particular, we find for the cases $Pr= \tau=0.1$ that $R_F \approx 1.5-2$, whereas \citet{2017ApJ...849...24M} found $R_F \approx 0.7$. 

\begin{figure}
\centering
\includegraphics[width=8cm]{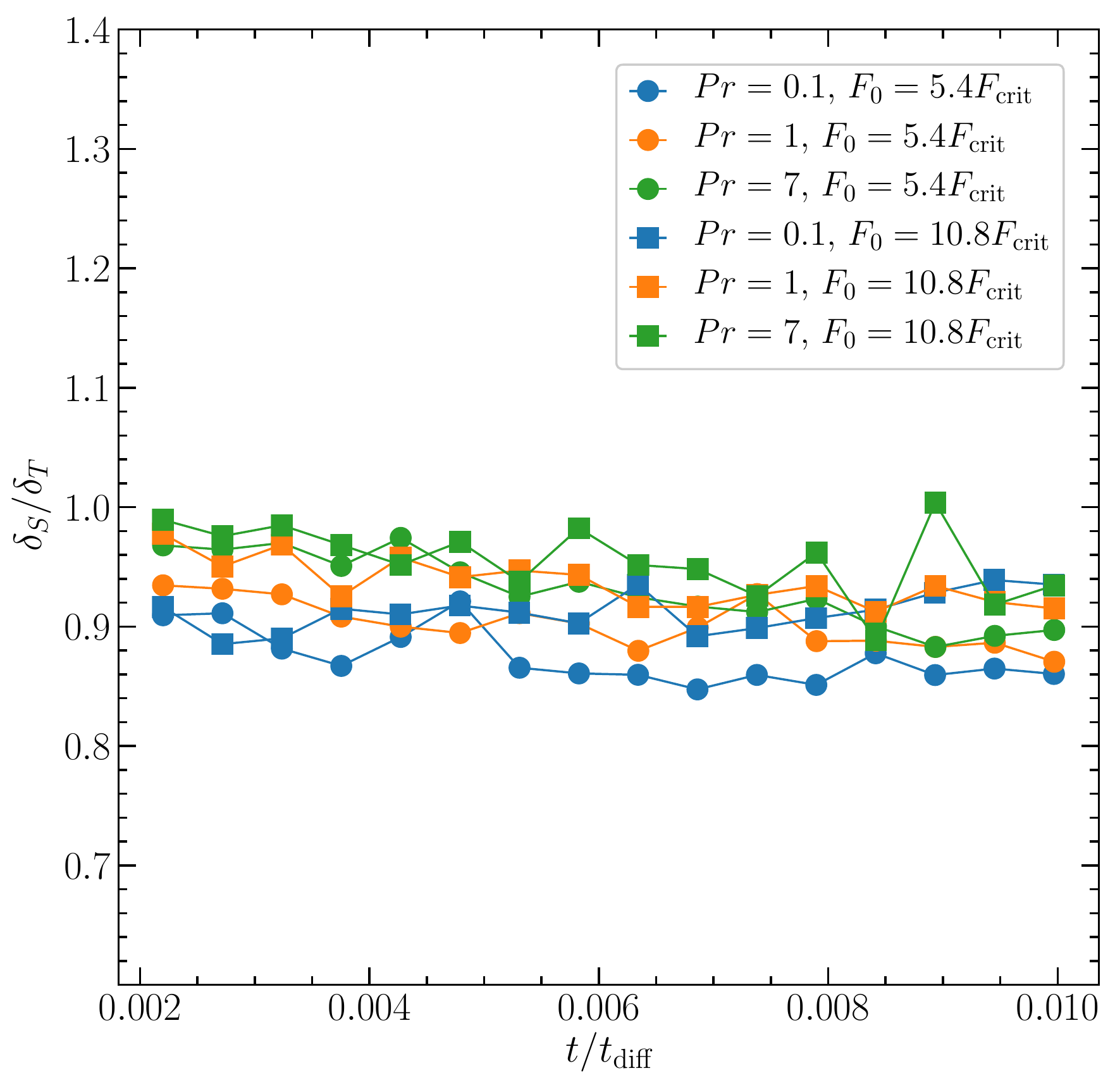}
\caption{Ratio between the thickness of the diffusive boundary layer of solute and temperature ($\delta_S/\delta_T$), as a function of $t/t_{\rm diff}$. As shown in the legends, colors distinguish between different $Pr$, and markers distinguish between different $F_0$.} \label{fig_thickness_ratio}
\end{figure}

Also shown in Fig.~\ref{fig_flux_ratio} are the values of $R_F$ computed using the diffusive fluxes of solute and heat only. In this case, the values are consistent with $R_F = \tau R_I$, as in the laboratory experiments by \citet{newell_1984}. As mentioned above, this implies that the diffusive boundary layers of solute and temperature have the same thickness. Indeed, direct measurement of the boundary layer thicknesses confirms this, and is shown in Fig.~\ref{fig_thickness_ratio}.

\section{Analytic Model for the Inwards Propagation of the Convective Layer} \label{sect_model}

The fact that the excess heat flux across the interface $\overline F_{\rm a}$ is a fixed fraction of the imposed heat flux, $\overline{F}_{\rm a} = \varepsilon F_0$ (section \ref{sec_flux_below}), and that the entrainment parameter $\gamma$ varies slowly in time (section \ref{sec_entrainment}), suggest the following set of equations to describe the location of the interface:
\begin{align}
h\dfrac{d\Delta \overline T}{dt} &= - \Delta \overline T \dfrac{dh}{dt} + \dfrac{F_0}{\rho_0 c_P}(1-\varepsilon)\, , \label{Eq_heat_balance}\\
\Delta \overline S &= \dfrac{1}{2}\left|\dfrac{d\overline S_0}{dz}\right|h\, , \label{Eq_salinity_balance} \\
{-\Delta \overline B}\, \dfrac{dh}{dt} &=  \gamma \left(\dfrac{ g\alpha F_{0}}{\rho_0 c_P}\right)\, \label{Eq_entrainment_balance},
\end{align}
where $\Delta \overline B = g\left(\alpha \Delta \overline T - \beta \Delta \overline S\right)$. This extends the analytic models of \citet{turner_1968} and \citet{fernando_1987} to include both entrainment and the heat flux across the interface.

It is worth noting that there is a separation of energy scales in this problem that allows us to write the global energy balance in Eq.~(\ref{Eq_heat_balance}) separately from the energy considerations that lead to the entrainment equation (\ref{Eq_entrainment_balance}). The energy required to mix the heavy elements, $E_{\rm mix} = \beta \rho_0 g \left|d\overline{S}_0/dz\right| H^3/12$ per unit area \citep{turner_1968}, is a small fraction of the total thermal energy lost by the layer,
\begin{equation}
\dfrac{E_{\rm mix}}{\rho_0 c_P\Delta \overline T H } = \dfrac{1}{6}\left(\dfrac{\beta\Delta \overline S}{\alpha \Delta \overline T}\right)\left(\dfrac{\alpha g H}{ c_P}\right)\sim \dfrac{\alpha g H}{c_P}\sim 10^{-7},\label{eq:energyfrac}
\end{equation}
where we write $\Delta \overline S = H\left|d\overline{S}_0/dz\right|/2$. Using mixing length estimates $F\sim \rho v_{\rm conv} c_P\delta T$ and $v_{\rm conv}^2\sim gH \alpha\delta T$ (where $v_{\rm conv}$ is a typical convective velocity and $\delta T$ a typical temperature fluctuation in the convection zone), we see that the kinetic energy flux $F_{\rm KE}$ associated with the convective motions is smaller than the thermal energy carried by convection by the same factor,
\begin{equation}
    \dfrac{F_{\rm KE}}{F}\sim \dfrac{\rho_0 v_{\rm conv}^3}{\rho_0 v_{\rm conv} c_P \delta T}\sim \dfrac{v_{\rm conv}^2}{c_P\delta T}  \sim \dfrac{\alpha gH}{c_P}. \label{eq:fluxfrac}
\end{equation}
Eq.~(\ref{Eq_entrainment_balance}) describes how this much smaller component of the energy, the kinetic energy,is used to entrain heavy fluid and move it across the interface. These contributions to the energy, however, are only small corrections to the overall thermal energy balance described by Eq.~(\ref{Eq_heat_balance}).

\begin{figure*}
\centering
\includegraphics[width=\textwidth]{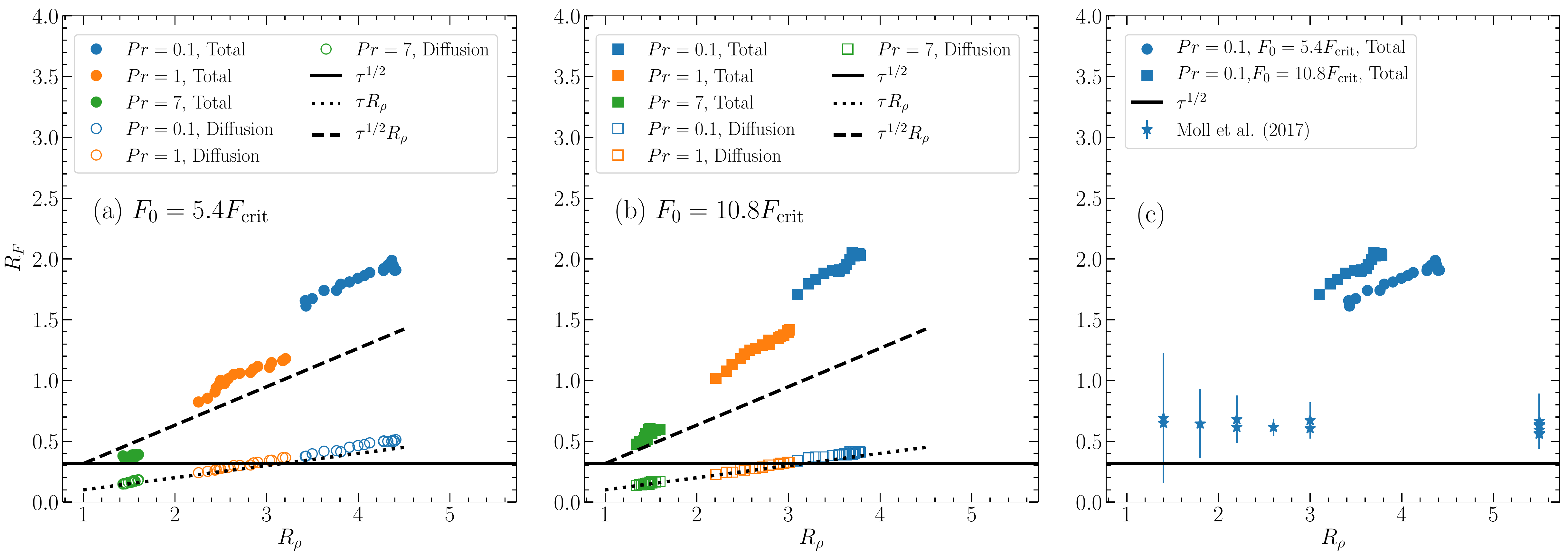}
\caption{Ratio between the buoyancy fluxes of solute and temperature, $R_F$, as a function of the density ratio parameter, $R_\rho$. Panels (a) and (b) show the results for $F_0 = 5.4 F_{\rm crit}$ and $F_0 = 10.8 F_{\rm crit}$, respectively. Panel (c) shows the results for the simulations at $Pr=\tau = 0.1$, and the results reported by \citet{2017ApJ...849...24M}. As shown in the legends, colors distinguish between different $Pr$, and markers distinguish between different $F_0$ as well as different flux ratios (filled markers consider the total flux, whereas unfilled markers consider just the diffusion flux). The black lines in all panels corresponds to different predictions (see text for more details).} \label{fig_flux_ratio}
\end{figure*}

\begin{figure*}
\centering
\includegraphics[width=\textwidth]{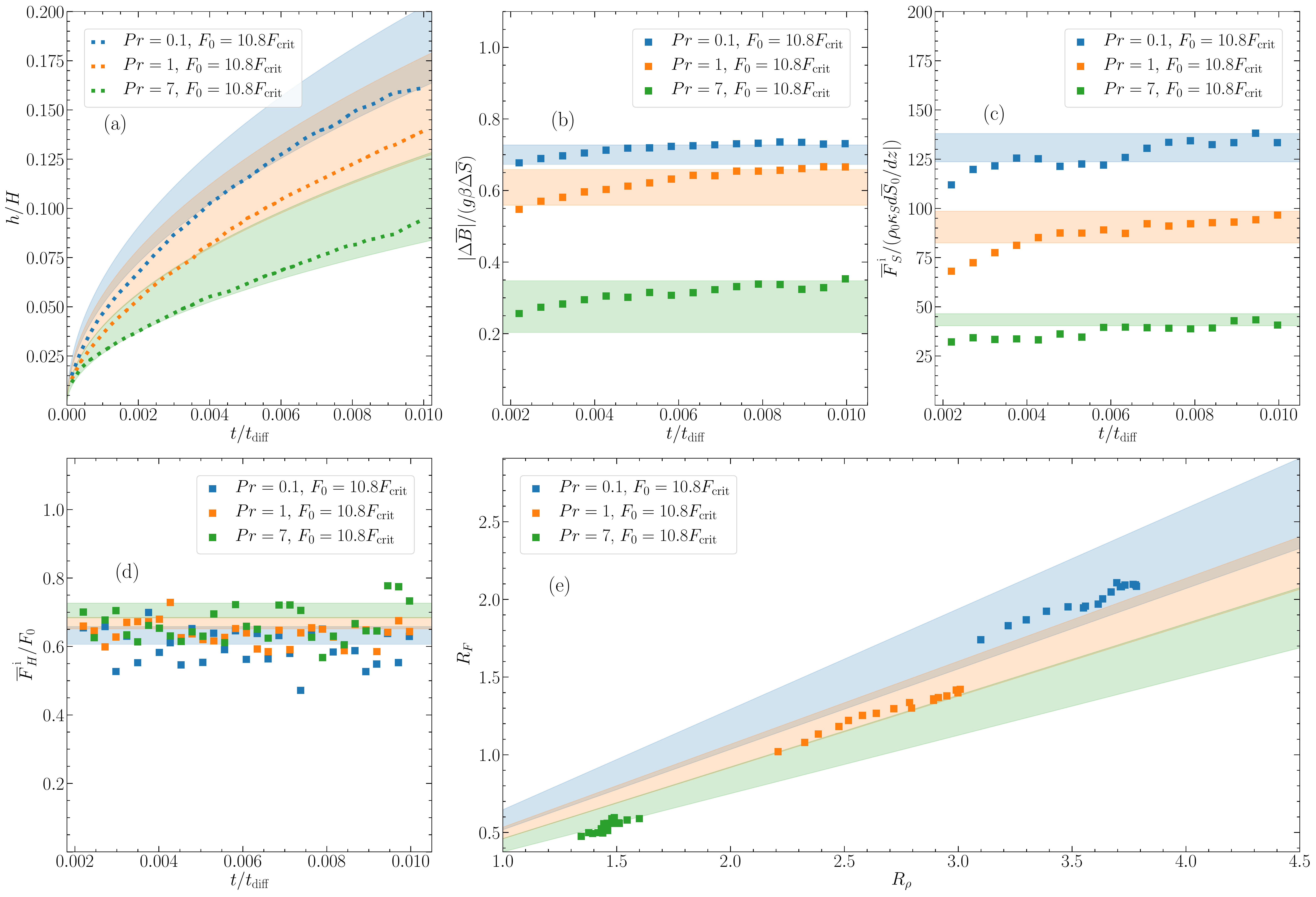}
\caption{A comparison of the analytic model given by Eqs.~(\ref{Eq_heat_balance})--(\ref{Eq_entrainment_balance}) with the results of our simulations. Panels (a), (b), (c), and (d) show for $Pr =$ (0.1, 1, 7), and $F_0=10.8F_{\rm crit}$, the temporal evolution (normalized to thermal diffusion time, $t_{\rm diff}$) of $h/H$, the ratio $|\Delta \overline B|/g \beta \Delta \overline S$, the solute flux through the interface normalized to the initial diffusive solute flux, $\overline F_S^{\, \rm i}/(\rho_0 \kappa_S |d\overline S_0/dz|)$, and the heat flux through the interface normalized to $F_0$, $\overline F_H^{\, \rm i}/F_0$, respectively. Panel (e) shows the relation between the buoyancy flux ratio, $R_F$, and the density ratio parameter, $R_\rho$. In all panels the shaded regions represent the predictions from the model by considering 1-sigma uncertainties in $\gamma$ and $\epsilon$. As shown in the legends, colors distinguish between different $Pr$} \label{fig_predictions}
\end{figure*} 

We now explore the consequences of this model. For simplicity and to get an analytic solution, we assume $\varepsilon$ and $\gamma$ constants (this choice is justified by the fact that both quantities vary slowly with time, with maximum variations at the level of less than 40\%). The set of equations (\ref{Eq_heat_balance})--(\ref{Eq_entrainment_balance}) has a solution $h\propto t^{1/2}$ which is
\begin{equation}
h(\gamma,\varepsilon,t) = \left(2C(\gamma,\varepsilon)\right)^{1/2} \left(\frac{F_0}{F_{\rm crit}}\right)^{1/2}\left(\kappa_T t\right)^{1/2},\label{eq_h_prediction}
\end{equation}
where $F_{\rm crit}$ is given by Eq. \eqref{eq_f_crit}. This is the same as Eq.~(\ref{eq_h_general}) but with a different prefactor. The constant $C$ is given in terms of the parameters $\gamma$ and $\varepsilon$ as
\begin{equation}
    C(\gamma,\varepsilon) = 1-\varepsilon + 2\gamma.
\end{equation}
Note that $\gamma$ and $\varepsilon$ can be measured directly from the simulations: see Fig.~\ref{fig_gamma_times} for $\gamma$ and Fig.~\ref{fig_f_h_in} for $\varepsilon$. For example, taking $\gamma\approx 0.85$ ($0.1$) and $\varepsilon\approx 0.3$ ($0.45$) gives $C \approx  2.4$ ($0.75$) for $Pr=0.1$ ($7$). Note that in the original model by \citet{turner_1968}, the constant $C$ is identified as $R_\rho$ which must be larger than unity (since an interface mixes by Rayleigh-Taylor instability as soon as it reaches $R_\rho=1$), so the fact that we infer $C=0.75$ for $Pr=7$ implies that additional physics must be at work.

We can also use Eqs.~(\ref{Eq_heat_balance})--(\ref{Eq_entrainment_balance}) and the solution Eq.~(\ref{eq_h_prediction}) to calculate the fluxes at the interface, and derive the expected relation between $R_F$ and $R_\rho$. First, Eqs.~\eqref{Eq_salinity_balance}, \eqref{Eq_entrainment_balance}, and \eqref{eq_h_prediction} give
\begin{align}
\dfrac{\Delta \overline B}{g \beta \Delta \overline S} & = - \dfrac{2\gamma}{1-\varepsilon + 2\gamma}\, , \\
 R_\rho = \dfrac{\beta \Delta \overline S}{\alpha \Delta \overline T} & =  \dfrac{1-\varepsilon + 2\gamma}{1-\varepsilon}.\label{eq_Rrho}
\end{align}
The first of these explains the ratio $\Delta \overline B/g \beta \Delta \overline S$ found in Fig.~\ref{fig_jumps} (panel d). To the extent that $\gamma$ and $\varepsilon$ vary slowly in time, so is the stability of the interface, which is determined by the values of $\gamma$ and $\varepsilon$. 
Again taking $\gamma\approx 0.85$ ($0.1$) and $\varepsilon\approx 0.3$ ($0.45$)  for $Pr=0.1$ ($7$), we find $R_\rho = 3.4$ ($1.4$) and $\Delta \overline B/g \beta \Delta \overline S = R_\rho^{-1} - 1.0 = -0.3$ ($-0.7$) (compare Figs.~\ref{fig_jumps} and \ref{fig_flux_ratio}). 

Eq.~(\ref{eq_Rrho}) shows that the range of values of $R_\rho$ depends on the maximum value of $\gamma$. The definition of $\gamma$ in eq.~(\ref{Eq_entrainment_balance}) suggests that $\gamma$ should not be much larger than unity, since in that case the energy required to mix fluid across the interface would exceed the available kinetic energy. With $\varepsilon = 0$, $R_\rho = 1 + 2\gamma$, which has a value $R_\rho=3$ when $\gamma=1$. This matches Turner's argument \cite{turner_1968} based on energetics for the maximum stability of the interface. When the heat flux across the interface is included, larger values of $R_\rho$ are possible, as seen in our simulations. For example, for the $Pr=0.1$ value $\varepsilon=0.45$, $R_\rho\approx 4.6$ for $\gamma=1$. The continued cooling of the convection zone continuously destabilizes the interface, preventing large values of $R_\rho$. 

The constant $C$ in eq.~(\ref{eq_h_prediction}) can be rewritten
\begin{equation}
    C(\gamma,\varepsilon) = R_\rho (1-\varepsilon),
\end{equation}
so we see that compared to Turner's estimate in eq.~(\ref{eq_h_prediction}), the height of the interface at a given time is smaller by a factor $(1-\varepsilon)^{1/2}$.

Eqs.~\eqref{Eq_heat_balance}, \eqref{Eq_salinity_balance}, and \eqref{eq_h_prediction}, also give expressions for the total flux of solute and heat through the interface
\begin{align}
& \overline F_{S}^{\, \rm i} = \rho_0 \Delta \overline S \dot{h} = \left(\dfrac{\alpha}{\beta}\right)\left(\dfrac{F_0}{c_P}\right)\left(\dfrac{1-\varepsilon + 2\gamma}{2}\right)\, , \\
&  \overline F_{H}^{\, \rm i} = \rho_0 c_P \Delta \overline T \dot{h} + \varepsilon F_0 = \dfrac{F_0}{2}\left(1+\varepsilon\right)\, .
\label{eq_fluxes_prediction}
\end{align}
The buoyancy flux ratio is
\begin{equation}
R_F = \dfrac{\beta \overline F_{S}^{\, \rm i}}{\alpha \overline F_{H}^{\, \rm i} c_P^{-1}} = \left(\dfrac{1-\varepsilon + 2\gamma}{1+\varepsilon}\right) = \left(\dfrac{1-\varepsilon}{1+\varepsilon}\right) R_\rho\, , \label{eq_buoyancy_flux_ratio_Rrho}
\end{equation}
which increases with $R_\rho$ as observed.

We compare the model predictions and the measurements from the simulations in more detail in Fig.~\eqref{fig_predictions}. By using the temporal averages and standard deviations of $\gamma$ and $\varepsilon$, we propagate their errors to get the uncertainties in the predictions above. We find that within the uncertainties, there is a good agreement between the model predictions and our numerical results. 

\section{Summary and conclusions} \label{sec_conclusions}

We studied the penetration of a cooling convection zone into a stably-stratified composition gradient at low $Pr$. Our goal was to extend previous work on salty water at Prandlt number $Pr\approx 7$ to low values $Pr<1$ found in planetary interiors. Our main conclusions are:

\begin{enumerate}

\item A non-negligible buoyancy jump develops over the interface between the convective layer and the stratified region (Fig.~\ref{fig_jumps}c). The stability of the interface as measured by the density ratio $R_\rho = \beta\Delta \overline S/\alpha \Delta \overline T$ increases slowly with with time as the convective layer grows, with a value ranging between $1\lesssim R_\rho\lesssim 4$ depending on Prandtl number.

\item Our results are well-described by an entrainment prescription in which a fixed fraction of the kinetic energy associated with the convective motions is used to lift heavier fluid across the interface, as proposed by \citet{fernando_1987} and \citet{molemaker_dijkstra_1997} for salty water. The entrainment efficiency $\gamma$ (Eq.~[\ref{eq_entrainment_balance}]) is approximately constant in time (with variations at the level of 20 - 40 $\%$). This confirms and extends to lower $Pr$ previous work identifying entrainment as the mixing mechanism responsible for the growth of the outer convective layer rather than Rayleigh-Taylor instabilities.

\item Entrainment is stronger at low $Pr$ and high imposed flux $F_0$. This implies that mixing is more efficient when the flow is more turbulent and energetic, with the result that the convective layer grows more quickly in those cases (Fig.~\ref{fig_thickness}). The entrainment parameter $\gamma$ changes from $\sim 0.1$ at $Pr=7$ to $\sim 0.9$ at $Pr=0.1$, so while entrainment is a relatively minor effect at $Pr=7$, it is much more significant at low $Pr$.

\item As pointed out previously by \citet{molemaker_dijkstra_1997}, additional interfacial heat flux, presumably associated with the transport of solute across the interface, is a significant fraction of the imposed heat flux at the top boundary (see Fig.~\ref{fig_f_h_in}). The flow of energy into the convective layer reduces the effective cooling rate of the convection zone.

\item We find that the interfacial heat and composition fluxes are dominated by advection rather than diffusion (this can be seen in Fig.~\ref{fig_flux_profiles}). Because the stability of the interface is limited to $R_\rho\lesssim 3$--$5$ (depending on $Pr$), it is always in a regime where advection dominates the interfacial transport. The interface adjusts so that the thickness of the temperature and salinity boundary layers are the same to $\approx 10$\%, despite the fact that the molecular diffusivities are different by a factor of ten ($\tau=\kappa_S/\kappa_T=0.1$). 

\item
Equations (\ref{Eq_heat_balance})--(\ref{Eq_entrainment_balance}) provide a simple analytical model that reproduces our numerical results with two parameters (assumed constant): the entrainment efficiency $\gamma$ (Fig.~ \ref{fig_gamma_times}), and the heat flux across the interface as a fraction of the applied heat flux at the top of the convection zone $\varepsilon=\overline{F}_{\rm a}/F_0$ (Fig.~\ref{fig_f_h_in}). The growth of the convection zone thickness follows $h\propto t^{1/2}$ and is given by Eq.~(\ref{eq_h_prediction}). Eq.~(\ref{eq_Rrho}) gives $R_\rho$ in terms of $\gamma$ and $\varepsilon$.

\end{enumerate}

Our focus in this paper has been on the growth of the outer convection zone, with the goal of addressing how low $Pr$ affects the rate at which it moves into the stably-stratified region. Another important question is whether secondary layers develop, slowing the progress of the convective region, and in principle preventing the system from mixing fully. Secondary layers are seen in salt water experiments, but it is not known when and how they arise in time-dependent cooling at low $Pr$. In this regard, a few attempts have been made \citep{2001PhDT.........8B,2019ThCFD..33..383Z}. \citet{2001PhDT.........8B} found that gravity waves can break near the interface and mix the composition gradient across, making the formation of secondary layers difficult to occur at low $Pr$. On the other hand, \citet{2019ThCFD..33..383Z} found that multiple layers can form at low $Pr$ either by a thermal instability at the interface ahead of the main convective layer, or spontaneously develop due to double-diffusive instabilities, as the ones observed in \citet{2003JFM...497..365R} and \citet{2012ApJ...750...61M}. We will discuss these issues in a companion paper.

We based our simulations on the pioneering salt-water experiments of \citet{1964PNAS...52...49T}, reducing the fluid viscosity to lower the Prandtl number. The lowest value of $Pr$ we consider, $Pr=0.1$, is at the upper end of values expected to occur in planetary interiors, where Prandlt numbers may extend down to $\sim 10^{-3}$. In stellar interiors, even lower values $Pr\sim 10^{-6}$ are expected. Our results suggest that the entrainment rate may be near maximum already at $Pr=0.1$, since $\gamma\sim 1$, implying that a large fraction of the available kinetic energy is taken up by entrainment. Recent calculations of convective boundary mixing in stars also find entrainment rates that scale linearly with the convective flux \cite{2015ApJ...798...49W,2017MNRAS.465.2991J,2020MNRAS.491..972A}, or in terms of bulk Richardson number as $\sim Ri^{-1}$ \citep{kato_phillips_1969,Meakin2007,Cristini2019}, supporting the kind of entrainment relation we have used here. An interesting difference is that in stars the composition difference is produced internally by nuclear burning and so the interface can be a lot stiffer than in our problem, where cooling of the convection zone quickly drives the stability of the interface $R_\rho$ to smaller values $R_\rho \lesssim 4$.

Even though entrainment at low $Pr$ involves a substantial fraction of the kinetic energy of convection, this energy is a small part of the overall energy budget (see Eqs.~[\ref{eq:energyfrac}-\ref{eq:fluxfrac}] and discussion in Sect.~\ref{sect_model}). The relevant energy is the kinetic energy because ultimately shear instabilities at the interface mix the fluid; the fact that buoyancy drives convective motions means that the kinetic energy is naturally of the same scale as the energy required to overcome the buoyancy of the stable interface.
This is important for core erosion in Jupiter: \citet{2017ApJ...849...24M} used the buoyancy flux ratio $R_F$ from their simulations to derive an expression for the core erosion rate that was substantially smaller than the earlier suggestion based on the total thermal flux integrated over the core radius \cite{2004jpsm.book...35G}. The ratio between the new erosion rate and the old rate is exactly the ratio (Eq.~[\ref{eq:fluxfrac}]) between the kinetic energy in convection and the thermal energy. The distinction between kinetic energy flux and heat flux is an important one in Boussinesq convection with $\alpha\ll 1$ (note that as in salty water $\alpha$ is also $\ll 1$ in Jupiter's interior \cite{2012ApJS..202....5F}). In stellar convection, the distinction is less important since there the equation of state is closer to ideal gas with $\alpha\sim 1$; still the kinetic energy flux can be as small as $\sim 0.01$ of the total heat flux (see discussion in \cite{Meakin2007}).

We have made a number of approximations which should be relaxed in future work. Although two-dimensional simulations have been successful at reproducing the classic laboratory experiments by \citet{turner_1965} and \citet{fernando_1987} (e.g.~see Appendix A of \cite{2019ThCFD..33..383Z}), it would be interesting to compare 3D simulations with the same setup with our 2D results, particularly at low $Pr$. Differences between 3D and 2D may explain the factor of $\approx 2$ lower values of $R_F$ measured at $Pr=0.1$ by \citet{2017ApJ...849...24M}, although their interface was between two convection zones rather than a convection zone and stable layer. In addition, in a planetary context, rotation and compressibility are important (see \cite{2017ApJ...834...44M} for a study of layer formation with rotation at low $Pr$), and so simulations that go beyond the Boussinesq approximation and include rotation would be of great interest.

\begin{acknowledgements}
We thank Pascale Garaud, Falk Herwig, Toby Wood, and Florian Zaussinger for useful discussions. This work was supported by  an NSERC Discovery Grant. We also thank
Ben A. and Evan Anders for technical support with the Dedalus code. J.R.F. acknowledges support from a McGill Space Institute (MSI) Fellowship. A. C. and J. R. F. are members of the Centre de Recherche en Astrophysique du Québec (CRAQ) and the Institut de recherche sur les exoplanètes (iREx). This research was enabled in part by support provided by Calcul Québec (calculquebec.ca), and Compute Canada (www.computecanada.ca). Computations were performed on Graham and Béluga.
\end{acknowledgements}

\bibliography{references}

\end{document}